\documentclass[11pt,reqno,tbtags]{amsart}
\usepackage[left=1in, right=1in, top=1in, bottom=1in]{geometry}
\usepackage{amssymb}
\usepackage{comment}
\usepackage{url} 
\usepackage{float}
\usepackage{mathrsfs}
\usepackage{graphicx} 
\usepackage{appendix}
\usepackage{placeins}
\usepackage{csquotes}
\usepackage{xcolor}
\usepackage{graphicx}
\usepackage{subcaption}
\numberwithin{equation}{section}

\newcommand{\sss}{\scriptscriptstyle}
\newcommand{\Top}[2]{\mathsf{Top}_{#2}^{G}\left(#1\right)}
\newcommand{\CCC}{\mathsf{CCC}^{G}_{R,S}}
\newcommand{\CCCrev}{\mathsf{CCC}^{G}_{S,R}}
\newcommand{\CCCo}{\mathsf{CCCo}^{G}_{R,S}}
\newcommand{\invisible}[1]{}
\newcommand{\eqn}[1]{\begin{equation} {#1} \end{equation}}

\usepackage{todonotes} %
\newcommand{\RvdH}[1]{\todo[inline, color=magenta]{Remco: #1}}

\newcommand{\ON}[1]{\todo[inline, color=purple]{Oliver: #1}}

\makeatletter
\newcommand*{\balancecolsandclearpage}{%
  \close@column@grid
  \cleardoublepage
  \twocolumngrid
}
\makeatother

\theoremstyle{definition}

\newtheorem{definition}{Definition}

\theoremstyle{remark}

\newcounter{thmenumerate}

\newcounter{xenumerate}

\begingroup
  \divide\count255 by 60
  \multiply\count255 by -60
  \advance\count255 by \time
  \ifnum \count255 < 10 \xdef\klockan{\the\count1.0\the\count255}
  \else\xdef\klockan{\the\count1.\the\count255}\fi
\endgroup

\def\rompar(#1){\textup(#1\textup)}    %

\def\xexp(#1){e^{#1}}

\renewcommand\phi{\varphi}

\def\[#1]{[\![#1]\!]}

\newcommand{\tempoff}[1]{#1}

\begin{document}

\title{Bringing Order to Network Centrality Measures}%
\author{G. Exarchakos}%
 \email{g.exarchakos@tue.nl}
 \address{%
Department of Electrical Engineering, Eindhoven University of Technology
}%
\author{R. van der Hofstad}
 \email{r.w.v.d.hofstad@tue.nl}
 \address{%
Department of Mathematics and Computer Science, Eindhoven University of Technology
}%

\author{O. Nagy}%
 \email{nagy@sdf.org}
\address{%
Department of Mathematics and Computer Science, Eindhoven University of Technology
}%

\author{M. Pandey}%
 \email{m.pandey@math.au.dk}
 \address{%
Department of Mathematics, Aarhus University
}%

\title{Bringing Order to Network Centrality Measures}

\date{}

\begin{abstract} 
We introduce a quantitative method to compare arbitrary pairs of graph centrality measures, based on the ordering of vertices induced by them. The proposed method is conceptually simple, mathematically elegant, and allows for a quantitative restatement of many conjectures that were previously cumbersome to formalize. Moreover, it produces an approximation scheme useful for network scientists. We explore some of these uses and formulate new conjectures that are of independent interest.
\end{abstract}

\maketitle

\invisible{\RvdH{Main TO DO's dd January 8, 2026:\\
$\rhd$ George will add a few potential use cases, and then we will consider to add a paragraph on this.
\begin{enumerate}
\item power grids: Identify critical components in Power Transmission Networks - often disagreements on which centrality best identifies “critical” nodes. Resilience analysis under failures and attacks - which nodes need protection given the cost of protecting to achieve cheapest maximum survivability.
\item communication networks: Backbone router and switch criticality... same as above focusing mostly at which network elements to protect.  
\item VLSI, SoC \& hardware architecture: which cores are bottlenecks and need redundancy paths. Do simple topological metrics identify the same hotspots as expensive simulation-based metrics? It can also be used for detecting which components age faster and which ones are more failure critical. We can compare age-aware centrality vs topological-based centrality: high CCC suggests structure-driven failure risks... low CCC means some physics need to be remodeled.
\item control systems:
\item CPS:
\item quantum: 
\end{enumerate}
}}

\section{Introduction}\label{sec:intro}
The importance of centrality measures in network science is similar to the importance of derivatives or integrals in calculus. Yet, while both concepts are \emph{intuitively} clear to practitioners, their mathematical details can lead to unexpected complications in general settings, thus making a single \emph{universally} applicable mathematical definition challenging. The case of centrality measures is arguably less straightforward, since there are many reasonable centrality measures that capture different aspects of what \enquote{being central} can mean in a network. To illustrate this point, the database \cite{centiserver} currently lists more than 400~centrality measures, each motivated by a different use case. Even very recently, new centrality measures are being introduced \cite{engsig2024domirank}. These various centrality measures are not only minor variations of the same idea; usually, there are no tractable relations between them. We refer to \cite[Chapter 7]{newman2018networks} and \cite{reviewcentrality2,reviewcentrality3} for a comprehensive introduction to centrality measures. 

In this paper, we propose a method to {\em directly} study how close centrality measures are in a particular network. This allows us to bring {\em order} to centrality measures by proposing a {\em quantitative method} to compare arbitrary pairs of centrality measures defined on the same graph. Our method is based on the \emph{vertex-ordering} induced by the pairs of centrality measures. Intuitively, it captures the disagreement between the {\em collections} of top-ranked vertices for two different centrality measures. We explain the method, discuss its properties, and discuss possible use cases.
\smallskip

\paragraph{\bf Background.} 
While there are some more light-hearted uses of centrality measures (e.g., the Kevin Bacon game \cite{NZ_herald, toronto_star} is related to closeness centrality, similarly, the Erd\H{o}s number \cite{Erdos_number} measures centrality in the collaboration network in mathematics), they have also been a subject of intense research  across disciplines. A classical topic is the resilience, fragility, and other structural properties of real-world networks \cite{Albert2000,CruaAsensio2017, bardoscia2021physics, battiston2012debtrank,boccaletti2006complex,Callaway2000,de2014navigability,Farooq2019,Gao2016,Guilbeault2021,jeong2001lethality}. Further, we specifically highlight uses in transportation networks \cite{Carvalho2009, Duan2014,Guimera2005}, epidemiology \cite{Kitsak2010,Pung2022,Salathe2010,Wang2016}, and fault-tolerance of power and telecommunication grids \cite{doyle2005robust,rinaldi2001identifying, Schafer2018,Sole2008}.

The {\em mathematical properties} of centrality measures have also received substantial attention. 
The PageRank centrality measure, introduced in \cite{BriPag98,page1999pagerank}, is said to follow the PageRank power-law hypothesis \cite{pagerankdistribution7, pagerankdistribution9,  pagerankdistribution1,GarHofLit20,pagerankdistribution3,MR4074702,pagerankdistribution5,pagerankdistribution6,MR2675117}, which states that in-degree and PageRank follow the same power-law distribution in most real-world networks  \cite{powerlawSF5,powerlawSF3,powerlawSF4,litvak2007degree,powerlawSF2}. Surprisingly, while often true, this hypothesis was recently disproved for directed preferential attachment models \cite{pagerankdistribution9}.

Centrality measures also attract interest in the physics community, since questions related to them can often be approached using methods from statistical physics. A recent example in this line of research is the use of the \emph{cavity method} in \cite{bartolucci2024distribution}. A recurrent topic in this line of research is the introduction of novel approximation schemes, such as \cite{bart1}, that allow practitioners to side-step the issues related to computational complexity of some centrality measures. We refer to \cite{SI2020} for an introductory overview of complexity and approximation issues.

{\em Comparisons} of centrality measures appear in \cite{boudin2013comparison, gaur2021comparison, ghazzali2017comparative, mocanu2018decentralized, vignery2020methodology}, often aiming to investigate which centrality measure outperforms others in a specific task. For example, \cite{mocanu2018decentralized} studies the effect of removal of top-ranked vertices according to various centrality measures on network connectivity, thus comparing centrality measures on their ability to identify connectivity hubs in real-world networks. Likewise, \cite{boudin2013comparison} compares the performance of various centrality measures in the specific task of keyword extraction from scientific articles. General comparisons of arbitrary centrality measures without any reference to a specific task are virtually non-existent, barring studies of correlations between centrality measures \cite{ghazzali2017comparative}. On a related note, there have been attempts to cluster centrality measures according to some similarities in their behaviour \cite{SI2020}, but this clustering is mostly heuristic and does not reflect any quantifiable clustering criteria. 

There is not much research on the {\em relations} between various centrality measures. In \cite{Evans2022}, an analytic derivation is given that relates closeness and degree centrality for undirected configuration models. Even though in centrality measures, we are mainly concerned with the induced rankings rather than the exact formula, it suggests that the inverse of closeness centrality of a vertex is linearly dependent on the logarithm of its degree. The dependence of PageRank on the damping factor is studied in \cite{boldi2005pagerank}.

{\em Quantifiable relations} between centrality measures can also be helpful for computational purposes. Indeed, many practical uses of centrality measures focus on the {\em ranking} of network vertices rather than the absolute value of the measure, and it is typically useful to identify the most central vertices, rather than to calculate the centrality of every vertex.
\smallskip

\paragraph{\bf Main innovation of this paper}
The main innovation of this paper is that we propose a way to {\em quantitatively compare} the rankings produced by different centrality measures. This is achieved by the so-called {\em Centrality Comparison Curve (CCC),} which describes the intersection of the top $x\cdot 100\%$ highest-ranked vertices in a network for any $x\in(0,1)$, and can be computed for any pair of centrality measures. From the CCC, one can immediately read off whether centrality measures are alike, or, instead, completely independent. Since the curve is computed for {\em every} $x\in (0,1),$ it illustrates where the centrality measures agree and where they are different. This is particularly useful when one wishes to compare the most central vertices in the network. 

We show that the CCC has several desirable properties. It is precisely the identity when two centrality measures are equal or are monotone transformations of each other, and it is insensitive to monotone transformations, which is desirable, since we are mostly interested in the \emph{ranks} of the centrality measures. Further, it is close to a square when the centrality measures are independent. Finally, we indicate that the CCC is well-behaved under large graph limits, both for local convergence of sparse graphs, as well as graphon limits for dense graphs. We showcase the use of the CCC by computing it for many pairs of centrality measures, both for artificial random networks as well as real-world networks, and investigate which pairs of centrality measures are close, close for the top-ranked vertices, and close to independent. In terms of quantifiable relations between centrality measures, our experiments indicate that different centrality measures can lead to similar rankings of the most central vertices. This has computational implications, since one can then substitute the calculation of a computationally complex measure by another of lower computational complexity. Thus, our work has the potential to significantly speed up the detection of the most central vertices for computationally hard centrality measures, but this is not the main focus of this paper.

We next define the {\em centrality comparison curve}.
\medskip 

\section{Results}\label{sec:results}

\subsection{The centrality comparison curve}\label{sec-ccc-definition}
Let $G=(V_G,E_G)$ be a graph with vertex set $V_G$ and edge set $E_G$. For simplicity, from now on we will assume that $V_G=\{1,\ldots, n\}=[n]$.

Recall that a {\em centrality measure} is a map $R\colon V_G\to {\mathbb R}_{\geq 0}$ such that $R(v)>R(u)$ should be interpreted as meaning that $v$ is more central than $u$. We can see $R$ as a naive way to order the vertices in $V_G$. For our purposes, we need to refine this ordering so that it becomes \emph{total} and it breaks ties consistently. To this end, we introduce the \emph{induced vertex ordering}: 

\begin{definition}[Induced vertex ordering]
\label{def:AIO}
    Fix a graph $G=(V_G,E_G)$, and let $R, S$ be two centrality measures on $G$. Endow each vertex $v\in V_G$ with a number~$u_v$ sampled independently from the standard $\mathrm{Unif}(0,1)$ distribution. Define the \emph{total} ordering $\prec_{(R, S)}$ such that, for any $a,b \in V_G$, $a\prec_{(R, S)} b$ if and only if
    \begin{enumerate}
        \item $R(a) < R(b)$; or
        \item $R(a) = R(b)$, and $S(a) < S(b)$; or
        \item $R(a) = R(b)$, $S(a) = S(b)$ and $u_a < u_b$.
    \end{enumerate}
    \hfill
\end{definition}
Note that $\prec_{(R, S)}$ is different from $\prec_{(S, R)}$. %
The total ordering in Definition \ref{def:AIO} allows us to define the main contribution of this paper, the Centrality Comparison Curve:

\begin{definition}[Centrality Comparison Curve]
\label{def-ccc}
Fix a graph $G=(V_G, E_G)$ with $n$ vertices. For any total ordering $\prec$ of $V_G$, denote the first $k$ elements in the ordered set $(V_G, \prec)$ by $\Top{k}{\prec}$. Take any two centrality measures $R, S$ defined over $G$, and consider the induced orderings $\prec_{(R, S)}$ and $\prec_{(S, R)}$ in Definition \ref{def:AIO}. For $A\subseteq V_G$, let $|A|$ denote the number of vertices in $A$.

The {\em centrality comparison curve} (CCC) on a graph of size $|V_G|=n$ is a mapping $\CCC \colon (0,1] \to [n]/n$ defined by
\begin{align}
\label{CCC-def}
    \CCC(x) = \frac{\big|\Top{\lceil xn\rceil }{\prec_{(R, S)}} \cap \Top{\lceil xn\rceil}{\prec_{(S, R)}}\big|}{n}.
\end{align}
In words, $\CCC(k/n)$ represents the overlap of the $k$ most central vertices according to the two centrality measures $R$ and $S$, normalised by the size of the graph.
\hfill
\end{definition}

In many applications, it is only relevant to what extent the two centrality measures agree on the set of most highly ranked vertices. For this purpose, we introduce a simple \emph{scalar} quantity derived from the CCC, given by
\begin{align}
    \CCCo(p) =\frac{\CCC(p)}{p},
\end{align}
which, for $p$ small, can be seen as an approximation of the right-derivative of $\CCC$ at $x=0$. Informally, one can expect roughly $n\CCCo(k/n)$ of the $k$ most highly ranked vertices for $R$ to also be in the $k$ most highly ranked vertices for $S$. For example, $\CCCo(0.05)$ measures how well the centrality measures $R, S$ agree w.r.t.\ the top 5\% of most central vertices. The closer it is to $1$, the better they agree.

The $\CCC$ makes no assumption about the edge density of the graph; hence it works for both sparse and dense graphs, as well as directed and undirected graphs, making it a universally applicable comparison.

\subsection{Properties of CCC }\label{sec-ccc-maths}
In this section, we discuss several properties of the Centrality Comparison Curve (CCC). 
\smallskip

\paragraph{\bf Symmetry of the CCC}
The CCC is symmetric in that $\CCC=\CCCrev$. 
\smallskip

\paragraph{\bf Monotonicity of the CCC}
Again by definition, $x\mapsto \CCC(x)$ is non-decreasing.
\smallskip

\paragraph{\bf Invariant under monotone increasing transformation}
In the literature, \emph{quantitative comparisons} of centrality measures are generally done by computing the {\em correlation} between centralities. However, correlation measures a {\em linear} dependence, and it can vary wildly when applying a monotone transformation. The advantage of the CCC is that it is {\em invariant} under monotone increasing transformations, and thus more robust, i.e., $\CCC$ is invariant under a simultaneous mapping of both $R$ and $S$ by strictly monotone increasing functions. As such, the CCC gives a quantitative way of comparing how close centrality measures are, as well as the most central vertices in the network, or, alternatively, how different they are.
\smallskip

\paragraph{\bf CCC is identity curve when $R=S$.} If $R = S$, then, by construction, no reordering occurs, hence
$\CCC(x) = x,$
which serves as the diagonal reference for perfect agreement. This fundamental property is lost under alternative tie-breaking rules, which is one reason for our tie-breaking rule. Also, it explains that how much $\CCC(x)$ is below $x$ is a measure of how {\em different} $R$ and $S$ are in their top $\lceil n x\rceil$ vertices with largest ranks.
\smallskip

\paragraph{\bf Bounded above by the identity curve}
The CCC always lies below the identity line, which is also a consequence of its definition, i.e., $\CCC(x)\leq x$ for all $x\in[0,1]$.  Thus, if $\CCC(x)$ is close to $x$, then $R$ and $S$ are approximately the same. Similarly, if $\CCC(x)$ is close to $x$ for small $x$, then $R$ and $S$ rank approximately the same vertices as being the most central.
\smallskip

\paragraph{\bf Lower bound for maximally different centrality measures}
Obviously, $\CCC(x)\geq 0$ for all $x\in[0,1]$, while $\CCC(x)\geq 2x-1$ for all $x\in[\tfrac{1}{2},1].$ Thus, pairs of centrality measures $R,S$ for which $\CCC(x)\approx \max\{0,2x-1\}$ can be considered to be {\em maximally different}.
\smallskip

\paragraph{\bf Using CCC for testing independence of Centrality Measures}
Further, observe that if $R$ is a {\em random} ordering of the vertices {\em independent} of everything else, then $x\mapsto \CCC(x)$ will converge to the function $x\mapsto x^2$ when the network size $n\rightarrow \infty$. Thus, when $\CCC(x)$ is close to $x^2$ for all $x\in[0,1]$, we can  think of the two centrality measures as being fully unrelated. 
\smallskip

\paragraph{\bf Summary of comparisons}
These observations give rise to natural comparisons: If $\CCC(x)$ is close to the identity graph, then $R,S$ induce very similar rankings, while these rankings are completely unrelated if $\CCC(x)$ is close to $x^2$, and even opposite when $\CCC(x)$ is close to $\max\{0,2x-1\}$.

\subsection{Putting the centrality comparison curve to practice}\label{sec-putting-ccc-to-practice}
We use CCC on artificial and real-world directed and undirected networks to see which centrality measures are alike, and which are not. 

For the artificial network, we use a directed configuration model \cite{Boll80b, MR1664335} of size $n=100,000$, with in- and out-degree power-law exponent~3. We run this computation 100 times with different samples of the configuration model, and indicate the standard deviation around the mean of the 100 runs with red error bars. 

A key real-world network considered is the directed \enquote{hep-ph} citation network \cite{SNAP}, one of the benchmark datasets in network science, which was first used in \cite{GehrkeGinspargKleinberg2003}. Other real-world networks that we have studied are the U.S.\ patent dataset maintained by the National Bureau of Economic Research and first studied in \cite{LKF2007}, and the Stanford web graph first studied in \cite{LesLanDasMah09}. The code can be found at \verb|https://github.com/nagyol/CCC|, while the full set of figures can be found at \verb|https://doi.org/10.5281/zenodo.18214457|.
Further details about these computations can be found in the supplementary material. 

We will explain the plots of CCCs on the example of PageRank and in-/out-degree.
\smallskip

\begin{figure}[ht]
\centering
\tempoff{\includegraphics[width=0.45\columnwidth]{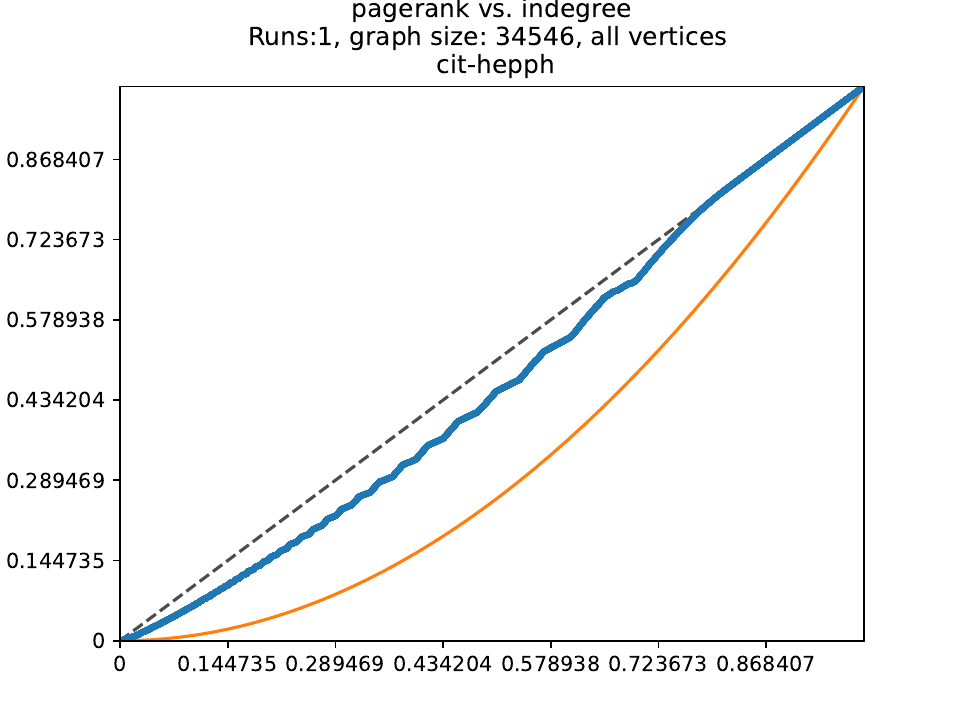}}
\tempoff{\includegraphics[width=0.45\columnwidth]{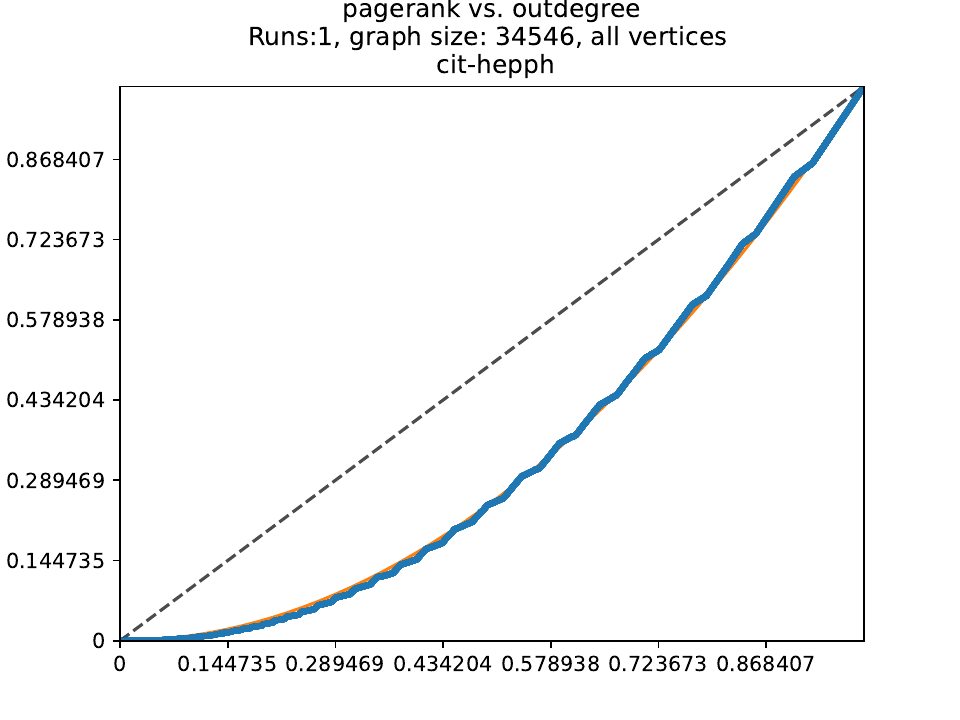}}
\caption{CCC for PageRank versus in-degree, and PageRank versus out-degree, for the \enquote{hep-ph} citation network \cite{GehrkeGinspargKleinberg2003}.}
\label{fig-citation-network-others}
\vspace{-0.3cm}
\end{figure}

\paragraph{\bf How to read a CCC plot.} We start by explaining the plots of CCCs on the example of PageRank versus in-/out-degree.
Figure \ref{fig-citation-network-others} shows that in the real-world citation network, PageRank is related to in-degree, and essentially independent of out-degree. The first conclusion follows from the fact that the blue CCC is reasonably close to the dashed identity curve in Figure \ref{fig-citation-network-others}(a), while the other follows since the CCC in Figure \ref{fig-citation-network-others}(b)  follows almost perfectly the yellow $x^2$ curve.

\vspace{-0.3cm}

\invisible{
\begin{itemize}
    \item figures, and their explanation. In paper, only one simulation. In supplementary material, also concentration one figure (rest in supplementary material)
    \item division into groups, name them? local ones (degree, pagerank, betweenness!!! + literature), global  ones, Katz?
    \item explain first for power-law CM, then argue that sth similar is there also for real-world network. 
\end{itemize}
}

\begin{figure}[H]
\centering
\tempoff{\includegraphics[width=0.43\textwidth]{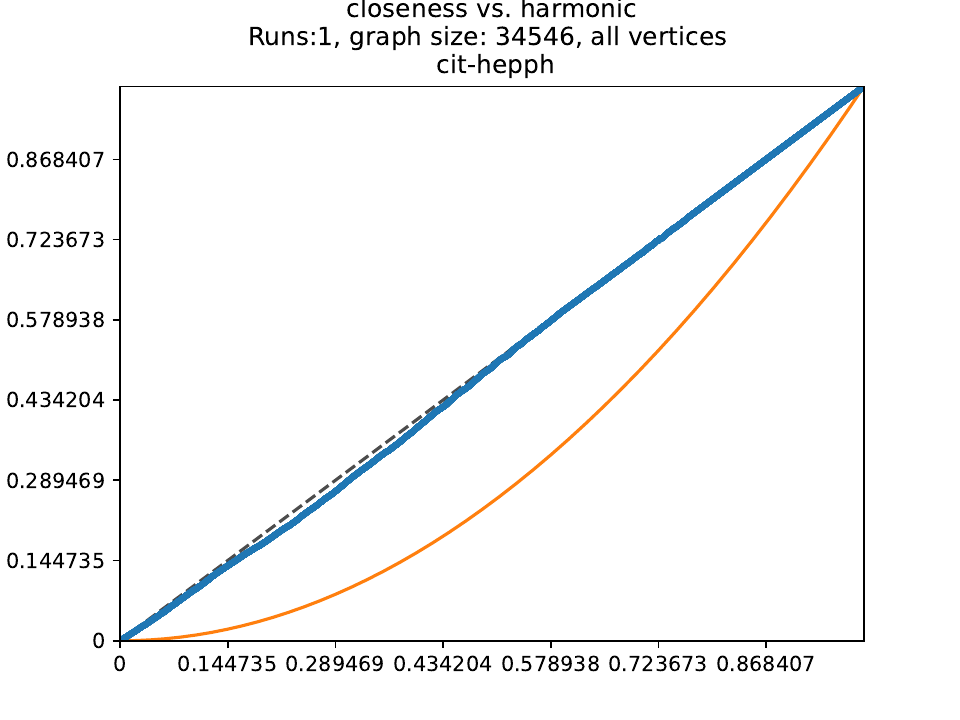}}
\tempoff{\includegraphics[width=0.43\textwidth]{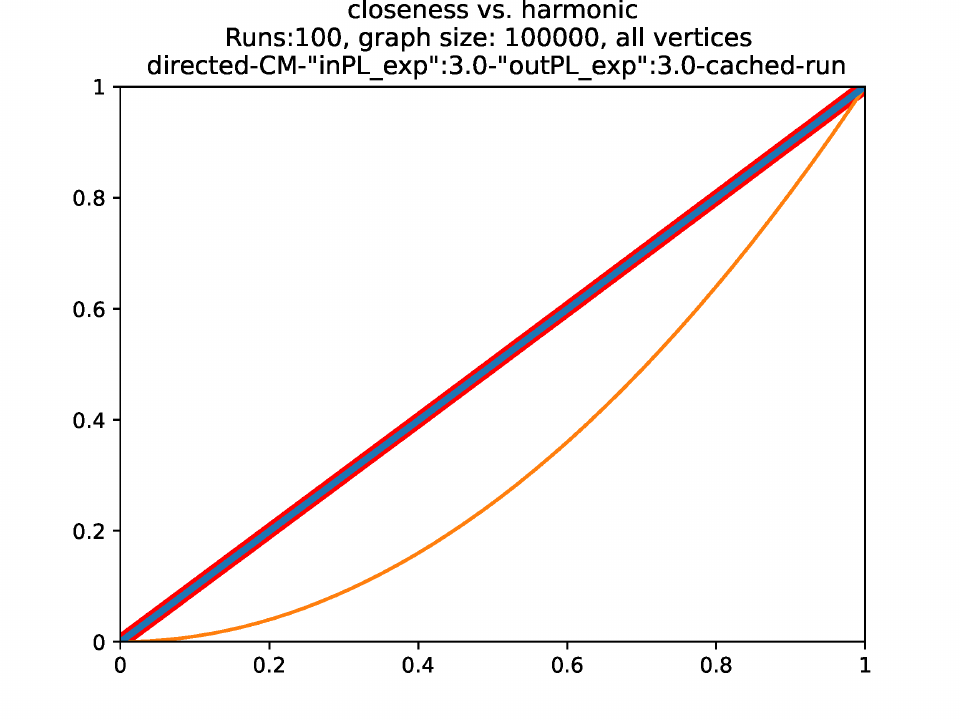}}
    \caption{CCC for closeness versus harmonic centrality, for the \enquote{hep-ph} citation network \cite{GehrkeGinspargKleinberg2003} and the artificial network.}
    \label{fig-closeness-harmonic}
\end{figure}

\begin{figure}[H]
\centering
\tempoff{\includegraphics[width=0.43\textwidth]{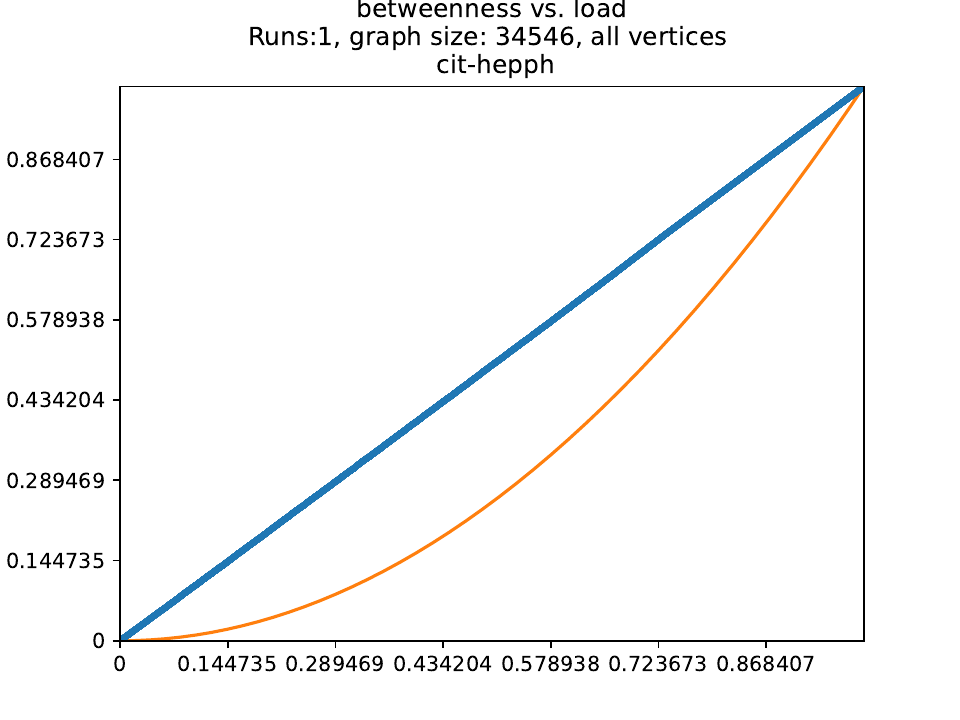}}
\tempoff{\includegraphics[width=0.43\textwidth]{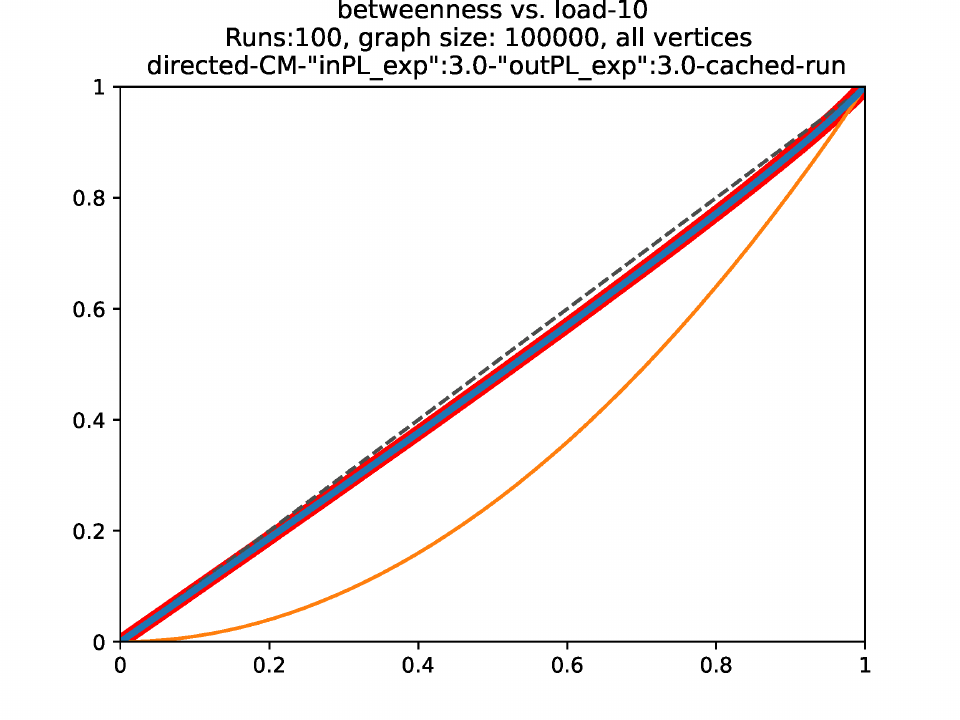}}
    \caption{CCC for betweenness versus load centrality, for the \enquote{hep-ph} citation network \cite{GehrkeGinspargKleinberg2003} and the artificial network.}
    \label{fig-betweenness-load}
\end{figure}

\smallskip
\paragraph{\bf Closeness and harmonic, and betweenness  and load, centrality are similar} As can be expected, due to the close relation in their definition, closeness and harmonic centrality are often close (see Figure~\ref{fig-closeness-harmonic}). A \emph{similar relation} holds for \emph{betweenness  and load centrality} (see Figure~\ref{fig-betweenness-load}).

\vspace{-0.3cm}
\vspace{-0.3cm}

\vspace{-0.3cm}
\vspace{-0.3cm}

\begin{figure}[H]
\centering
\tempoff{\includegraphics[width=0.45\textwidth]{figures/pagerank-indegree-hep.pdf}}
\tempoff{\includegraphics[width=0.45\textwidth]{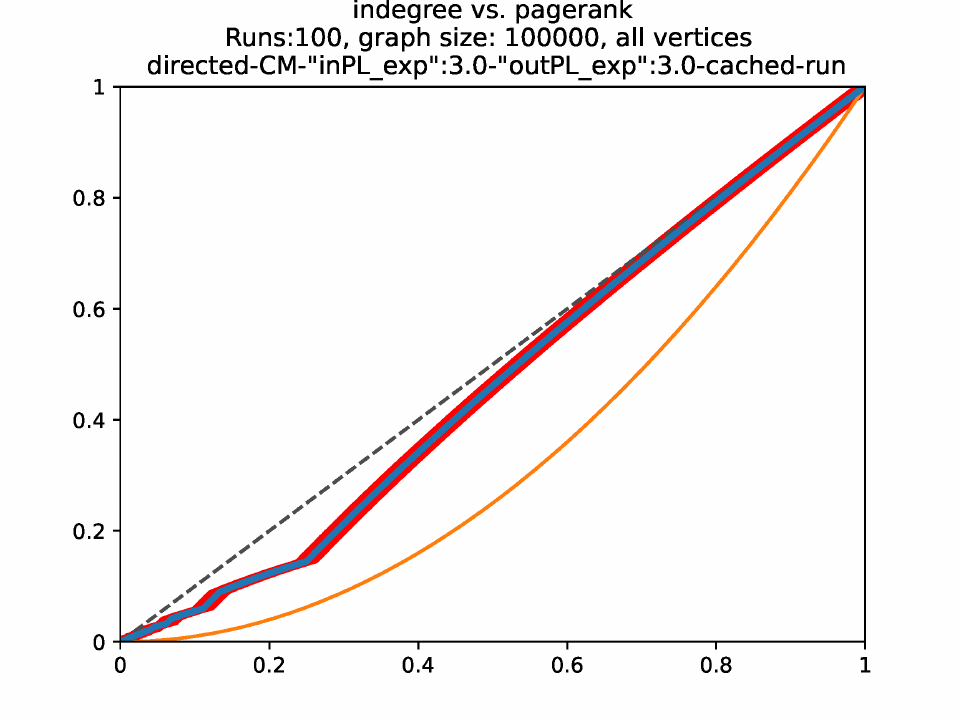}}
    \caption{CCC for PageRank versus in-degree, for the \enquote{hep-ph} citation network \cite{GehrkeGinspargKleinberg2003} and the artificial network.}
    \label{fig-PageRank-degree}
\vspace{-0.5cm}
\end{figure}

\smallskip
\paragraph{\bf PageRank weakly follows in-degree}
While in-degree centrality is simply the in-degree of any given vertex, PageRank is a more involved stochastic process originally used for ranking pages in Google search results \cite{page1999pagerank}. There is a large body of work showing that the PageRank distribution has the same power-law exponent as the in-degree distribution \cite{pagerankdistribution9,powerlawSF5, chen2017generalized,powerlawSF3,powerlawSF4,MR4074702,litvak2007degree, powerlawSF2}. As a result, unsurprisingly, the CCC for these centralities is reasonably close to a straight line 
(see Figure~ \ref{fig-PageRank-degree}).

\vspace{-0.3cm}
\noindent

\begin{figure}[ht]
\centering
\tempoff{\includegraphics[width=0.45\textwidth]{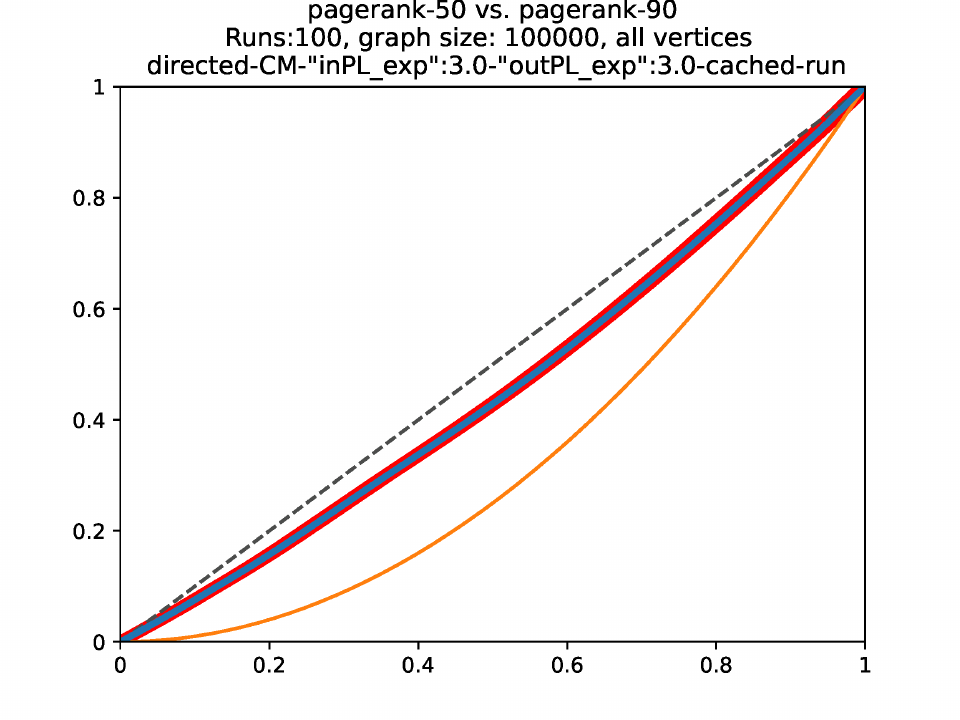}}
\tempoff{\includegraphics[width=0.45\textwidth]{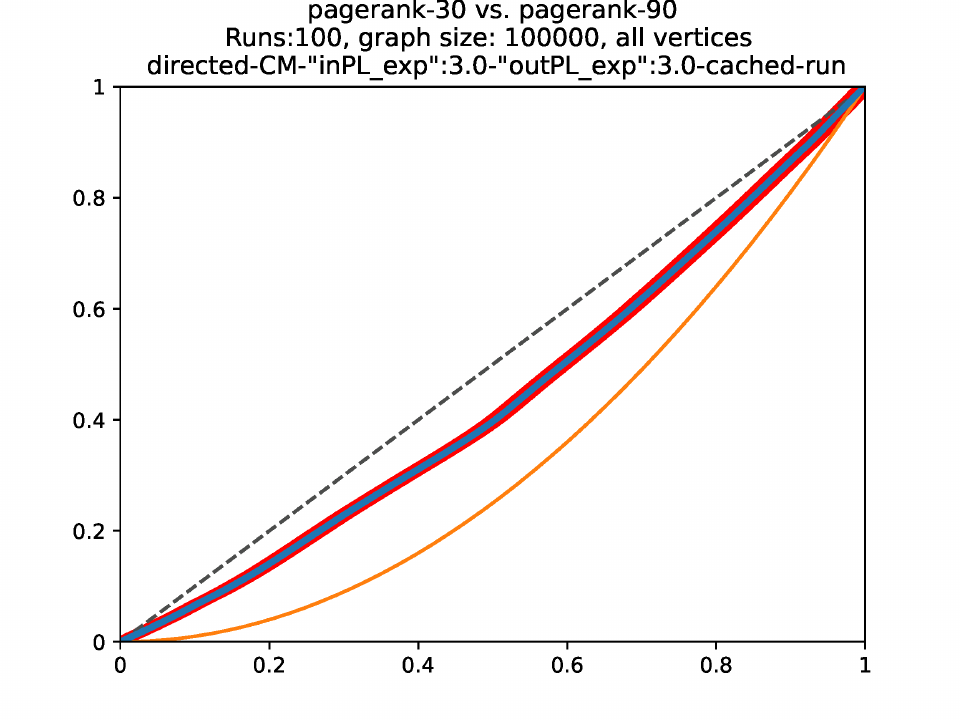}}
    \caption{CCC for PageRank with damping factors 0.5 versus 0.9, and 0.3 versus 0.9, respectively, for the artificial network.}
    \label{fig-PageRank-damping}
\vspace{-0.5cm}
\end{figure}

\smallskip
\paragraph{\bf PageRank hardly depends on damping factor} PageRank centrality has a parameter, called the {\em damping factor}, that is often taken as 0.85 \cite{page1999pagerank}. However, the precise choice of the damping factor does not significantly matter, as can be seen in Figure~\ref{fig-PageRank-damping}.

\vspace{-0.3cm}
\noindent

\invisible{\smallskip
\paragraph{\bf Betweenness and degree agree on top vertices}
Betweenness and degree centralities differ noticeably. However, in the artificial network, there is a significant overlap among the top-ranked vertices, whereas this overlap is less evident in the collaboration graph (see Figure~\ref{fig-betweenness-degree}).

\begin{figure}[ht]
\centering
\tempoff{\includegraphics[width=0.4\textwidth]{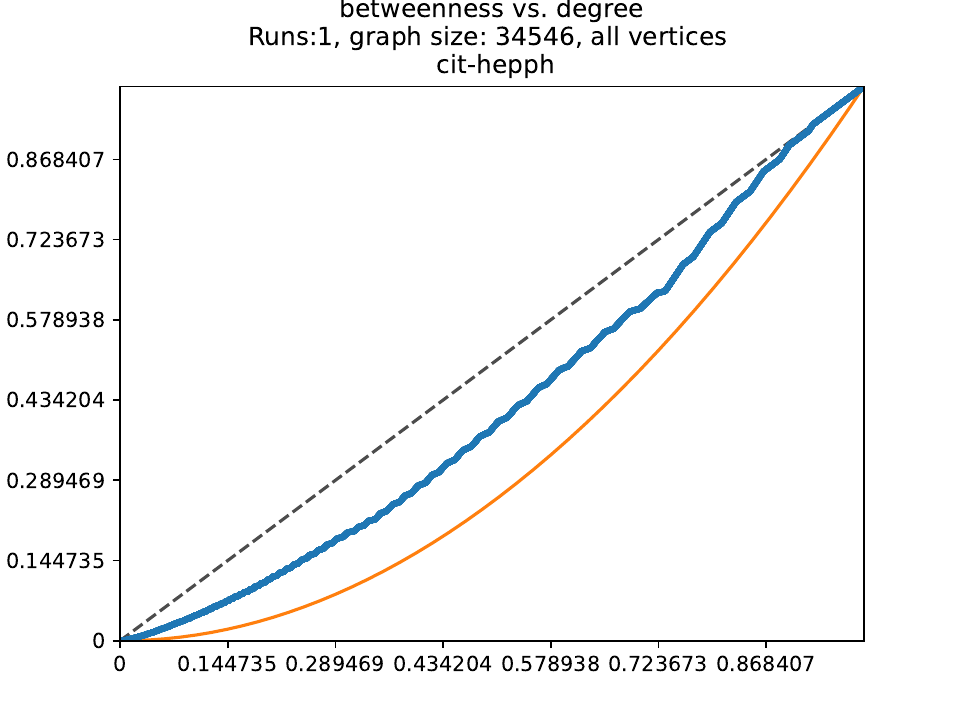}}
\tempoff{\includegraphics[width=0.4\textwidth]{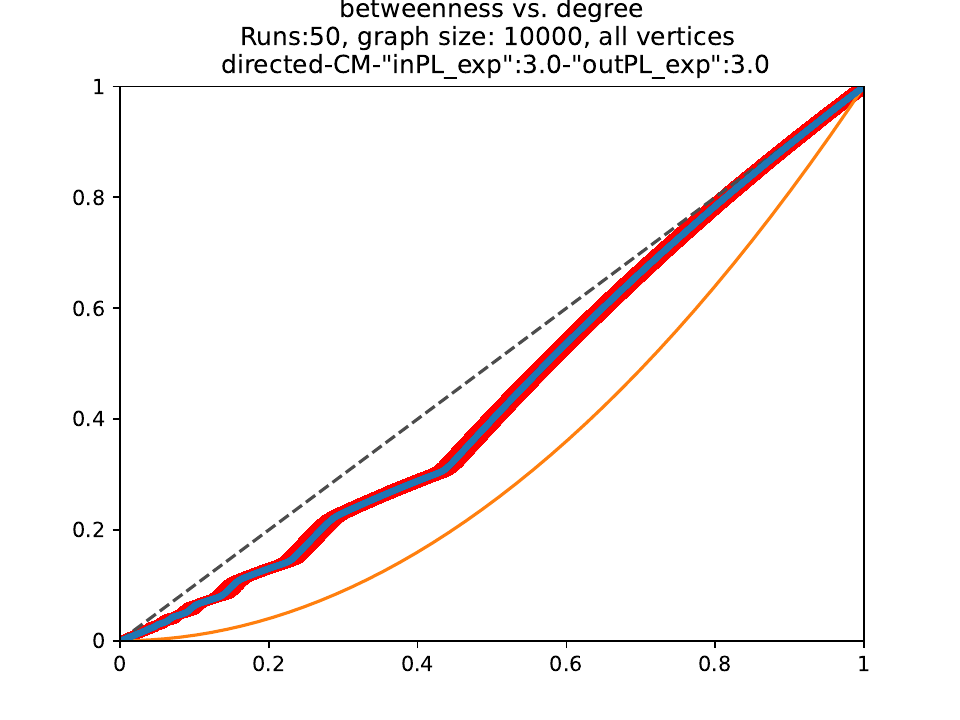}}
    \caption{CCC for betweenness and degree, for the \enquote{hep-ph} citation network \cite{GehrkeGinspargKleinberg2003} and the artificial network.}
    \label{fig-betweenness-degree}
\end{figure}

\noindent}

\begin{figure}[h!]
\centering
\tempoff{\includegraphics[width=0.32\textwidth]{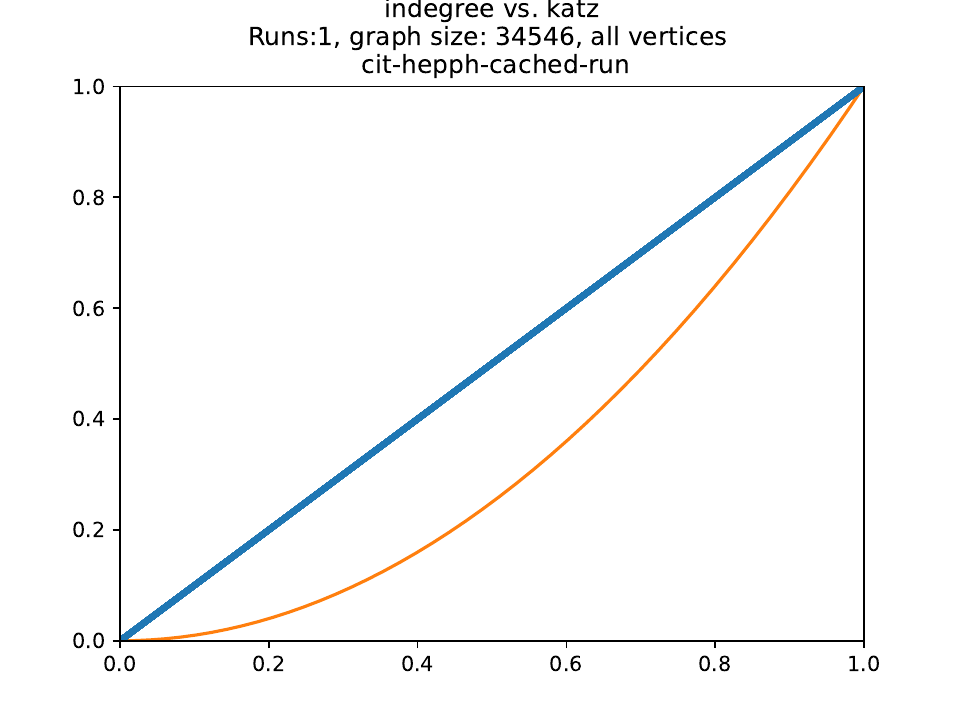}}
\tempoff{\includegraphics[width=0.32\textwidth]{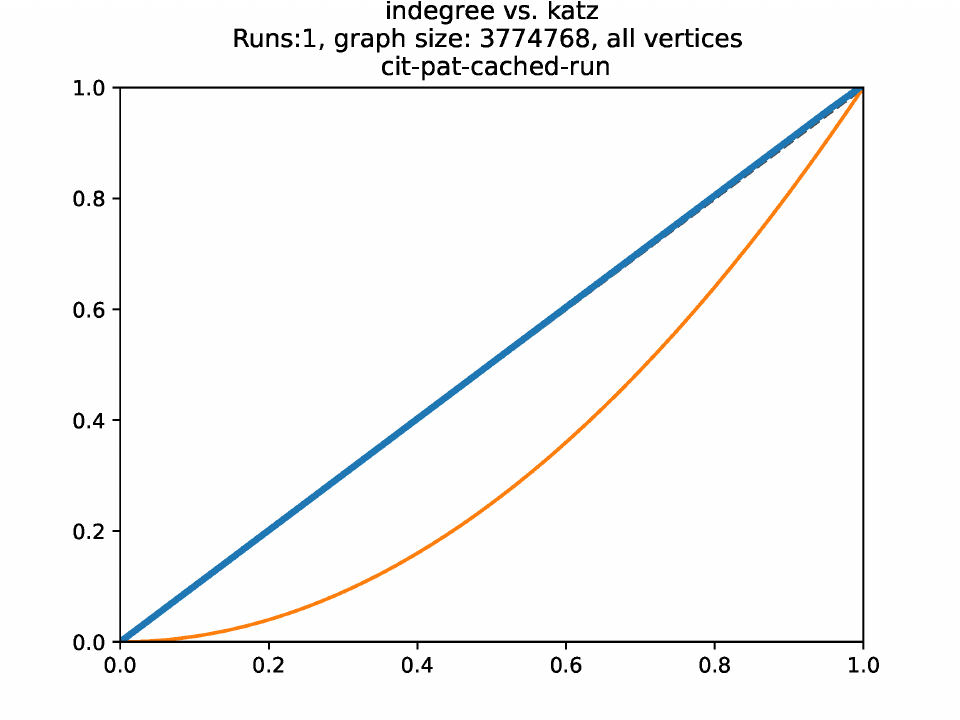}}
\tempoff{\includegraphics[width=0.32\textwidth]{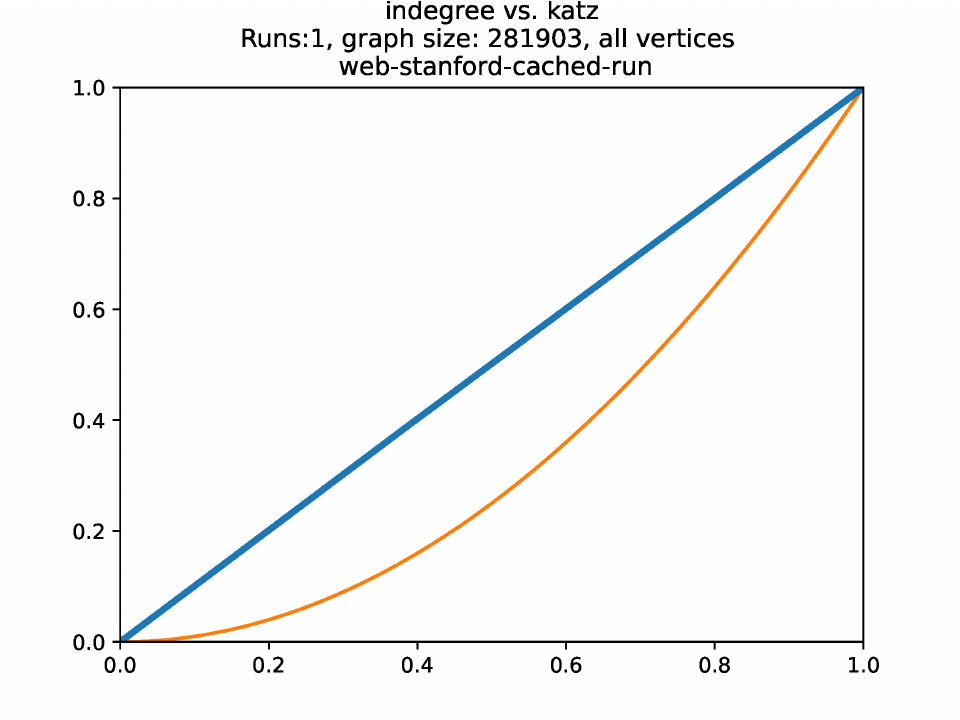}}
\caption{CCC for in-degree versus Katz centrality for the \enquote{hep-ph} citation network \cite{GehrkeGinspargKleinberg2003}, the \enquote{Patent} network \cite{LKF2007}, and Stanford web graph dataset \cite{LesLanDasMah09}.}
\label{fig-indegree-Katz}
\end{figure}

\smallskip
\paragraph{\bf Katz is similar to in-degree}
Interestingly, Katz centrality and in-degree are almost identical across the real-world datasets considered (see Figure~\ref{fig-indegree-Katz}).

\begin{figure}[h!]
\centering
\tempoff{\includegraphics[width=0.45\textwidth]{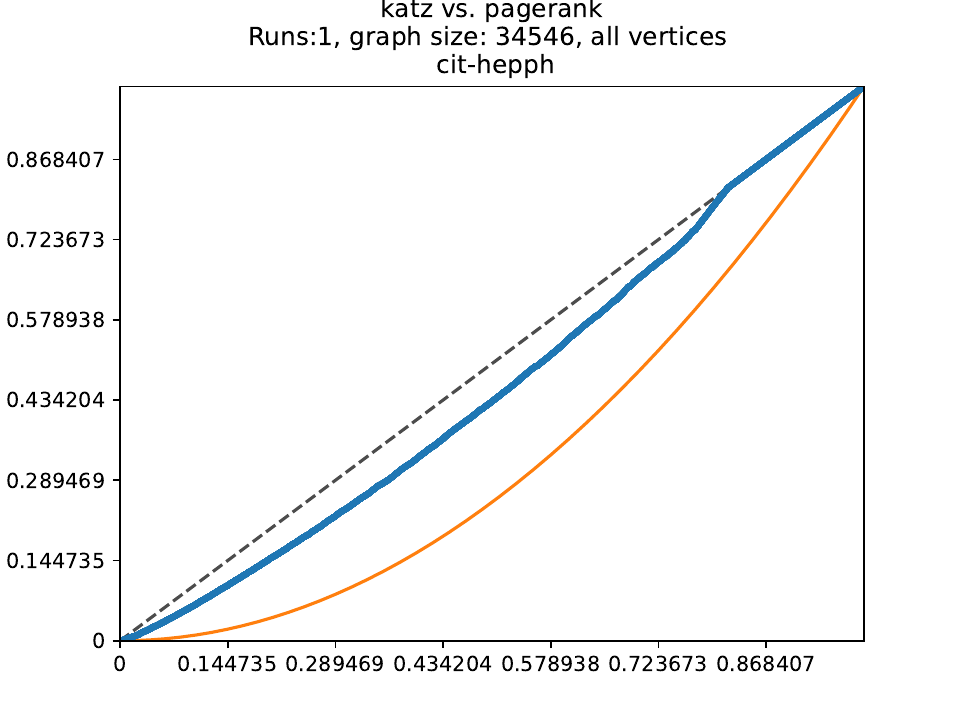}}
\tempoff{\includegraphics[width=0.45\textwidth]{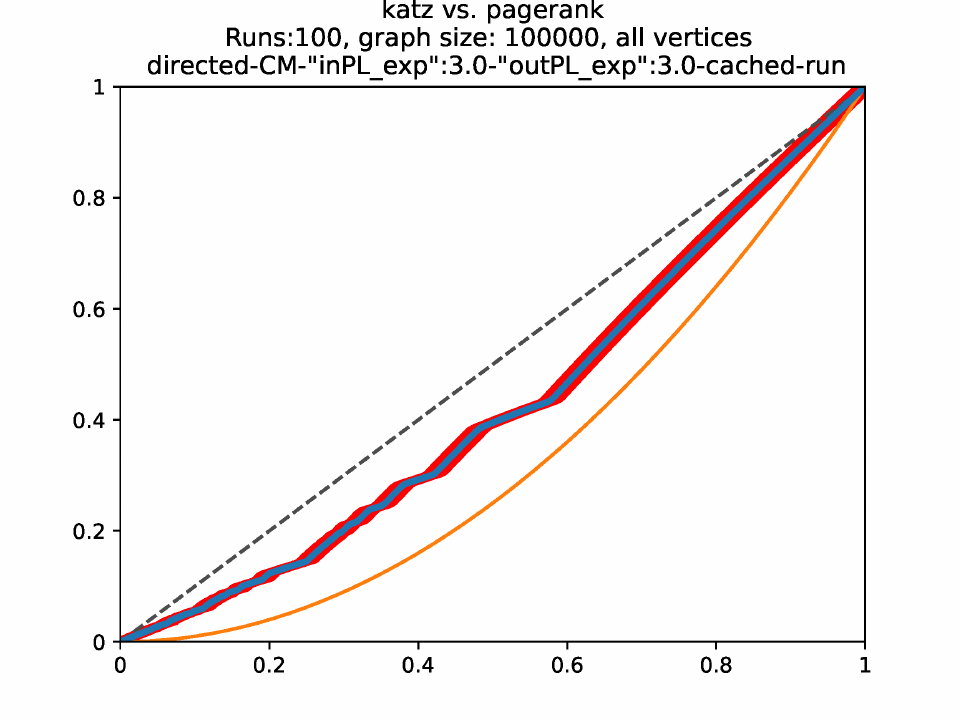}}
    \caption{CCC for PageRank versus Katz, for the \enquote{hep-ph} citation network \cite{GehrkeGinspargKleinberg2003} and the artificial network.}
    \label{fig-PageRank-Katz}
\end{figure}

\smallskip
\paragraph{\bf PageRank is somewhat different from Katz} Perhaps surprisingly, even though the definitions of PageRank and Katz centrality are highly similar, their centralities behave somewhat differently (see Figure~\ref{fig-PageRank-Katz}).

\begin{figure}[H]
\centering
\tempoff{\includegraphics[width=0.45\textwidth]{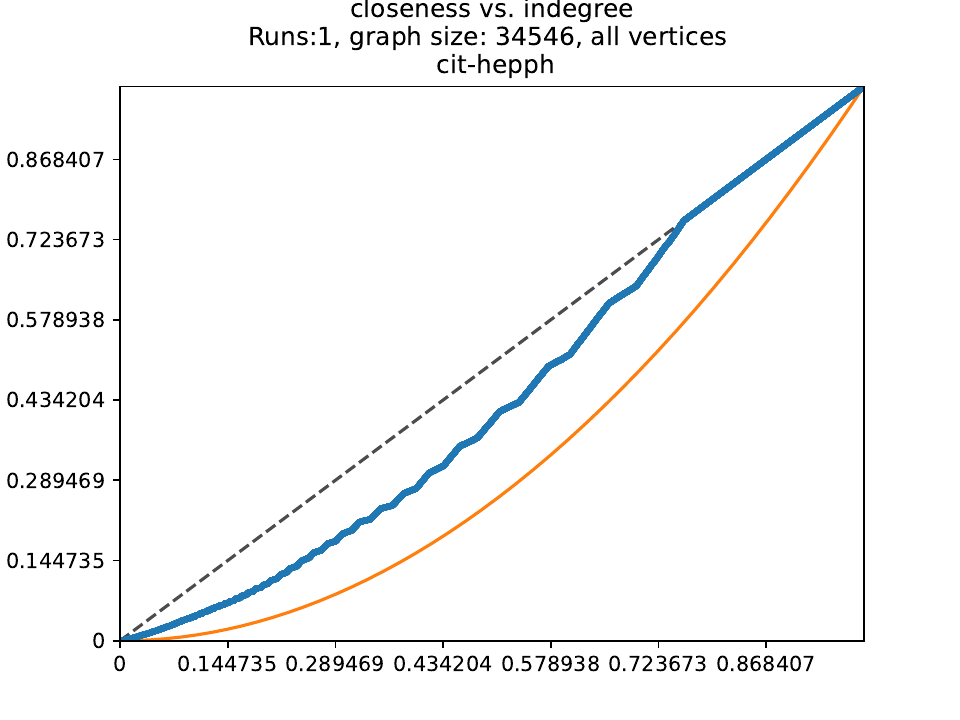}}
    \tempoff{\includegraphics[width=0.45\textwidth]{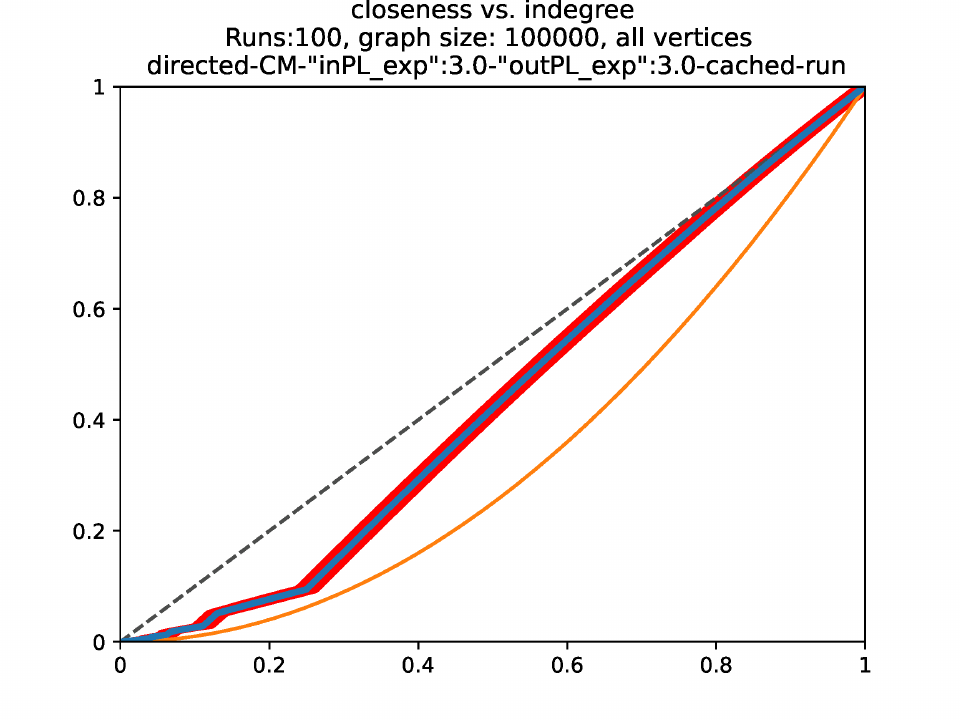}}
    \caption{CCC for closeness versus in-degree, for the \enquote{hep-ph} citation network \cite{GehrkeGinspargKleinberg2003} and the artificial network.}
    \label{fig-closeness-degree}
\end{figure}

\smallskip
\paragraph{\bf Closeness and in-degree are weakly related} In the directed configuration model, closeness and in-degree are somewhat related (see Figure~\ref{fig-closeness-degree}).

\smallskip

\begin{figure}[H]
\centering
\tempoff{\includegraphics[width=0.45\textwidth]{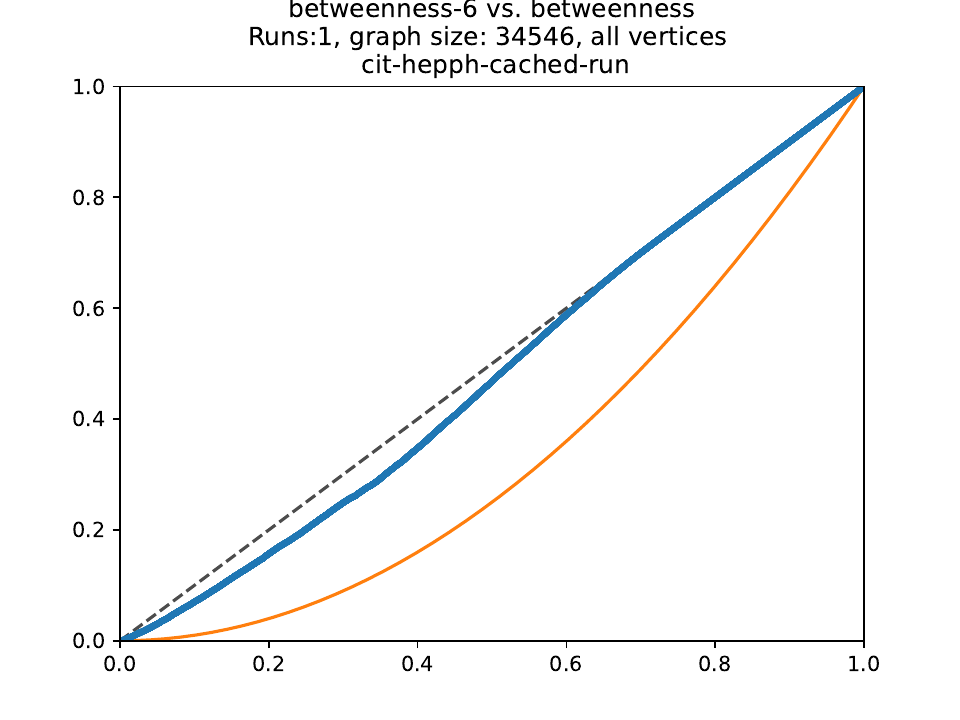}}
\tempoff{\includegraphics[width=0.45\textwidth]{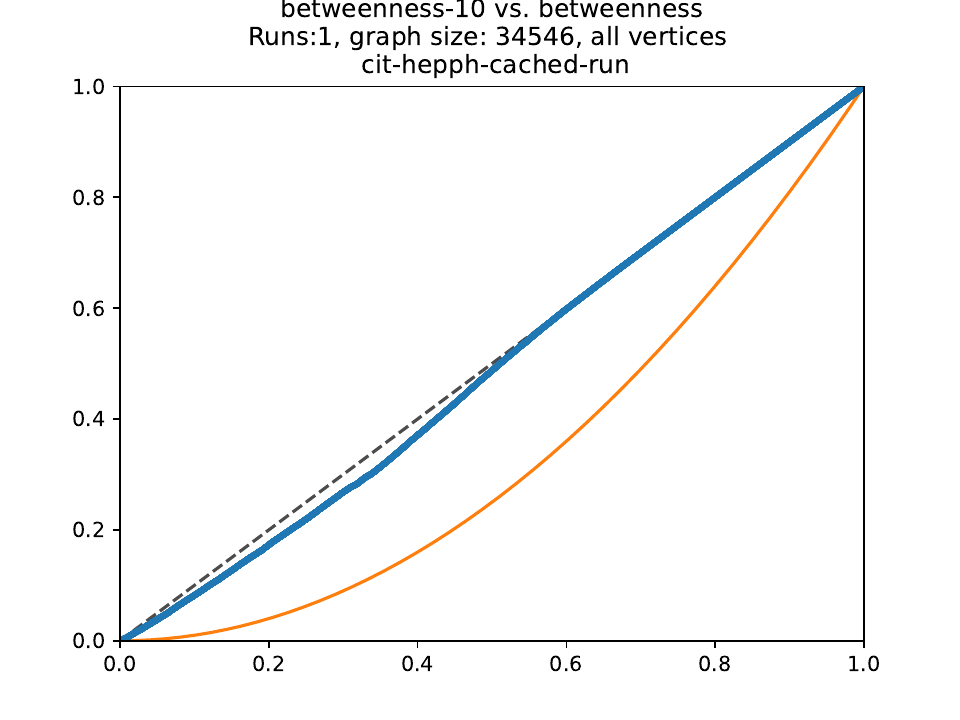}}
    \caption{CCC for betweenness versus betweenness6 and betweenness10, respectively, for the \enquote{hep-ph} citation network \cite{GehrkeGinspargKleinberg2003}.}
    \label{fig-local-betweenness}
\end{figure}

\begin{figure}[H]
\centering
\tempoff{\includegraphics[width=0.45\textwidth]{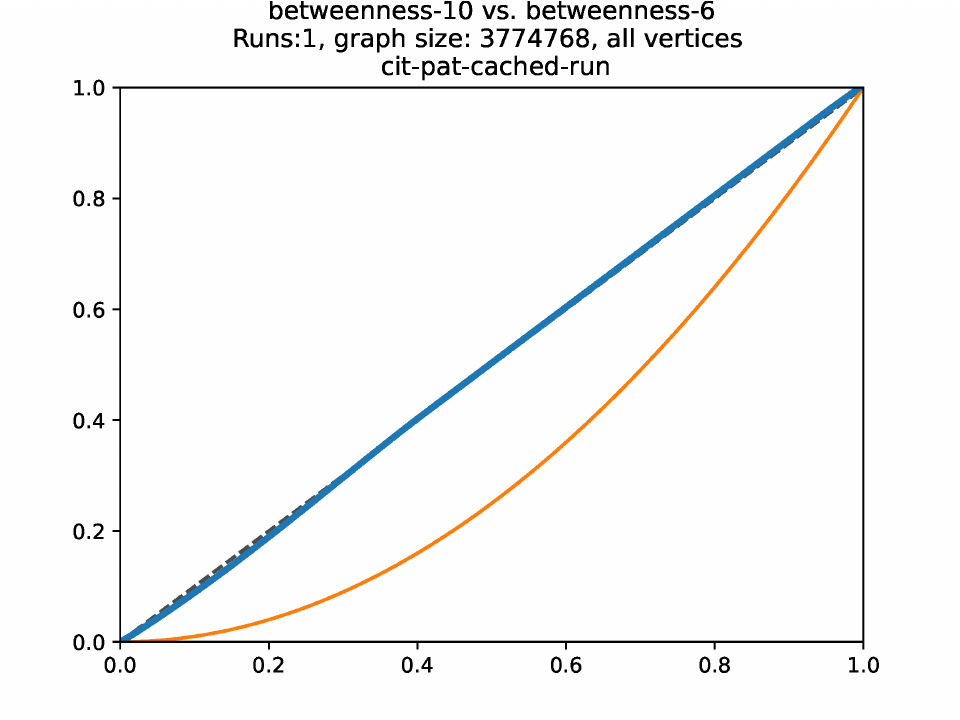}}
\tempoff{\includegraphics[width=0.45\textwidth]{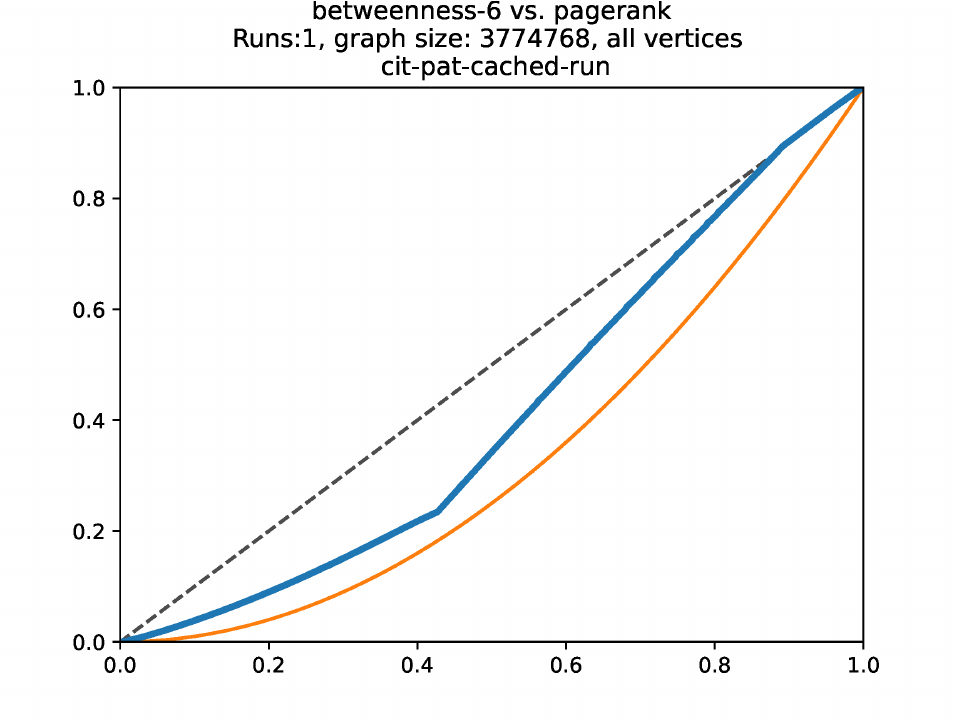}}
    \caption{CCC for  betweenness6 versus betweenness10, and betweenness6 versus PageRank, respectively, for the \enquote{Patent} network \cite{LKF2007}.}
    \label{fig-local-betweenness-pat}
\end{figure}

\paragraph{\bf Computational efficiency using the CCC} The CCC can be used to simplify the computation of computationally hard centrality measures, such as betweenness centrality. This is exemplified by Figure \ref{fig-indegree-Katz}, since in-degree is computationally significantly easier than Katz centrality, yet, for these real-world networks, they are virtually identical. For another example, Figure \ref{fig-local-betweenness} compares betweenness centrality with betweenness centrality based on paths of length at most 6 and 10, denoted here as betweenness6 and betweenness10, respectively, showing excellent agreement, certainly in the most highly ranked vertices. This allows one to approximate betweenness centrality in settings where it is computationally out of reach, by a restricted version that is computationally easier. For example, Figure \ref{fig-local-betweenness-pat} compares betweenness6 and betweenness10, showing excellent agreement, as well as betweenness10 and PageRank. For this large data set, of 3,774,768 vertices and  	16,518,948 edges, betweenness itself is computationally quite heavy.

\subsection{Convergence of centrality measures and the CCC in the large-graph limit}
\label{sec-convergence-CCC}
In this section, we discuss convergence of the CCC with respect to the two most important convergence notions of graphs sequences: {\em local convergence} \cite{AldSte04, BenSch01} for sparse graph sequences, and {\em graphon convergence} \cite{Lova12,Zhao23} for dense graph sequences. 
\smallskip

\paragraph{\bf Convergence of centrality measures and the CCC for local centrality measures} Local convergence \cite{AldSte04, BenSch01}, and locally tree-like behaviour, have become central techniques in network science. PageRank is known to be a {\em local} quantity, in that the PageRank of a vertex  can be well-approximated by PageRank on a finite-radius neighbourhood \cite{GarHofLit20}. Obviously also in-degree, out-degree and degree centrality satisfy this.  Due to its definition, one can also expect Katz centrality to be a local measure. 

For pairs of (jointly) local centrality measures, one can show that the CCC converges to a limiting continuous curve \cite{WIP}, 
that can be determined by the local limit of the graph. However, it is {\em not} true that local measures are all alike (see, for example, Figure~\ref{fig-PageRank-Katz}), even though their {\em relations} can be analysed through the local limit. At the same time, many {\em global} centrality measures, such as closeness, betweenness and load centrality, cannot be expected to converge for sparse graph sequences, as they are not defined by local neighbourhoods. Possibly, though, they may sometimes be {\em well-approximated} by local approximations, such as can be observed for betweenness in Figures \ref{fig-local-betweenness}--\ref{fig-local-betweenness-pat}. 

The fact that the CCC for some pairs of centrality measures converges for locally convergent graph sequences is illustrated by the fact that the CCCs for our artificial networks, which are random graph sequences that are known to converge locally, show little variation for different simulations. Indeed, the variability over different random graph simulations is indicated by the red lines in the CCCs in the figures, and these appear to be varying rather little. In turn, this is an important property, as it shows that the CCC is {\em robust} with respect to minor variations in the graphs in question, which is desirable. An example of how this could be used is to replace the {\em global} betweenness centrality measure by the {\em local} betweenness$k$ based on paths of length $k$ only. See Figures \ref{fig-local-betweenness} and \ref{fig-local-betweenness-pat} for examples of how this can be effectively used.
\smallskip

\paragraph{\bf Convergence of centrality measures and the CCC for dense graph sequences} In sparse graphs, the number of edges grows linearly in the network size. For {\em dense} graphs, instead, the number of edges grows {\em quadratically} in the number of vertices. For such graph sequences, {\em graphon convergence} is the appropriate limiting notion \cite{Lova12,Zhao23}. In many cases of convergent dense graph sequence convergence, closeness, betweenness and load centrality converge as well, so that also the CCC converges for pairs of such centrality measures. This can be shown, for example, for dense graphs with independent edges and {\em strictly} positive graphon limits. We refer to \cite{liu2025graphon} for an example of the use of graphon limits of the Katz centrality measure, and its implications for production networks.
On the other hand, for centrality measures such as (in-, out-)degree centrality, the {\em order} of the degrees per vertex are to a large extent determined by the {\em degree fluctuations}, which are not captured by the graphon limit. Such problems may be alleviated when assigning appropriate continuous {\em edge weights}, and computing the centrality measures for such weighted graph sequences.

\section{Discussion}
\label{sec-discussion}
In this section, we discuss our main results and pose open problems.

\smallskip
\paragraph{\bf Properties of the CCC} 
The Centrality Comparison Curve  has many desirable properties, in that it depends on {\em centrality ranking} only, and thus is insensitive to monotone transformations. It is the identity curve {\em precisely} when two centrality measures agree, and equals $x\mapsto x^2$ when two centrality measures are completely independent. Finally, since it is a {\em curve} that records the agreement for the top $x\cdot 100\%$ highest ranked vertices of both centrality measures, it also allows one to deduce for which rank cohorts the measures disagree. In particular, it allows for settings where two centrality measures are {\em rather different} for most of the vertices, yet {\em agree almost completely} for the most highly ranked vertices. Often, this is what matters most for practitioners. As such, it yields an attractive {\em quantitative measure} to compare pairs of centrality measures.
\smallskip

\paragraph{\bf Conclusions from the simulations}\label{sec-conclusion-simulations}
The CCC reveals that load centrality is closely aligned with betweenness centrality, as well as closeness centrality with harmonic centrality, which is consistent with their respective definitions. Additionally, PageRank is similar to in-degree centrality and remains largely unaffected by the choice of damping factor, while, despite their similar definitions, PageRank and Katz centralities are not as close as one might expect. 

In the undirected case, eigenvector centrality and PageRank, which also share similar definitions, behave almost like independent centrality measures. Similarly, degree and Katz centrality in the undirected collaboration network appear to act as independent centrality measures. See the supplementary material for the relevant figures.

Knowing which centrality measures are similar has numerous applications, for example in that one only needs to compute the algorithmically cheapest of similar centrality measures. It also allows one to test whether an intuitive ranking, such as produced by field experts, is well aligned with any of the standard centrality measures. Finally, in settings of {\em dynamic} networks, one can see how the centralities change over time by plotting the CCCs comparing different time pairs.

In the literature, centrality measures are generally compared by computing the {\em correlation} between centralities. However, correlation measures a {\em linear} dependence, and it can vary wildly when applying a monotone transformation. The advantage of the CCC is that it is {\em invariant} under monotone transformations, and thus more robust. As such, the CCC gives a quantitative way of comparing how close centrality measures are, as well as the most central vertices in the network, or, alternatively, how independent they are.
\smallskip

\paragraph{\bf Conclusion for convergence of the CCC for convergent graph sequences} The fact that the CCC has nice convergence properties for the two central graph convergence notions is rather promising, and indicates the usefulness of the CCC. At the same time, more research is needed to unleash the full potential of the convergence of the CCC.
\smallskip

\paragraph{\bf Open problems for the CCC}
There are many open problems for comparing centrality measures through the centrality comparison curve, and here we list two. 

The first would be to do a {\em large-scale investigation} of the relation between centrality measures for a large number of real-world networks. This was done for the tail behaviour of the degree distribution by Broido and Clauset \cite{BroCla19}, as well as in \cite{VoiHooHofKri19}, to investigate whether scale-free networks are abundant, or rather rare, based on the KONECT database \cite{Kune13}. The fact that these papers come to rather different conclusions exemplifies the fact that such an analysis is challenging. For centrality measures, it would involve to quantify when centrality measures are {\em close}, both for the entire set of vertices, as well as for the most highly ranked vertices, which is already non-trivial. It would also be useful to quantify what it means for centrality measures to be independent. Then, one would need to decide whether the various pairs of centrality measures are close, far, or independent. This should be done for {\em each network instance}. The final step would be to summarize these conclusions over the large body of networks to come to a general understanding of which pairs of centrality measures are similar, which are different, and which are independent. 

The second open problem relates to {\em rigorously establishing convergence properties} of the CCC, for sparse random graphs, for dense random graphs, and for everything in between. Again, this problem has several facets. One could study the convergence for {\em specific random graph models}, such as the popular network models, and investigate the similarity of pairs of centrality measures on them. Secondly, and more in line with the types of results that we have discussed in Section \ref{sec-convergence-CCC}, it is of central importance to prove the convergence for {\em general} random graph models in the large-graph limit. The fact that there are examples for which the CCC behaves well under {\em both} large-graph limits, local for sparse and graphon for dense, is a promising sign that merits further investigation.

\section{Methods: CCC and Graph limits}

\subsection{CCC and total order}

Centrality scores typically induce only a \emph{partial order} due to {\em ties} originating from vertices having {\em equal rank}. The \emph{Centrality Comparison Curve (CCC)} is rank-based, and thus works best when using a \emph{total order}. To obtain such a total order, we introduce a structured tie-breaking rule that preserves interpretability of the CCC.

Suppose $(R, S)$ denote the two centrality measures under consideration. 
In defining the CCC (recall Definition~\ref{def:AIO}), vertices are first ranked according to the primary measure~$R$. 
Tie-breaking then proceeds according to the following steps:
\begin{enumerate}
    \item Ties in $R$ are first resolved using the secondary measure~$S$.
    \item Any remaining ties are broken by independent $\mathrm{Uniform}(0,1)$ random variables assigned to each vertex.
\end{enumerate}
The resulting total order is denoted by~$\prec_{(R,S)}$,
and, analogously, initiating the ranking from~$S$ produces the order~$\prec_{(S,R)}$.
\smallskip

The {\em rationale} behind the two tie-breaking rules is that
Step~(1) ensures that ties in~$\prec_{(R,S)}$ are resolved using~$S$ (and symmetrically for~$\prec_{(S,R)}$), 
thereby \emph{maximizing the overlap} of the top-$k$ vertex sets in the two rankings. 
This guarantees that the CCC reflects genuine differences between $R$ and $S$ rather than artifacts of tie-breaking.
Then, Step~(2) ensures that the ranking really is \emph{total}, since $\mathrm{Uniform}(0,1)$ variables come from a continuous distribution and thus break any remaining ties with probability one.

\begin{figure}[h!]
    \centering
    \tempoff{\includegraphics[width=0.45\textwidth]{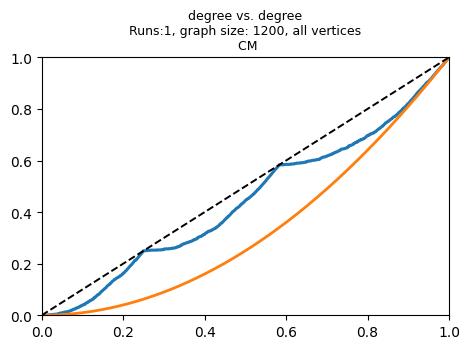}}
\tempoff{\includegraphics[width=0.45\textwidth]{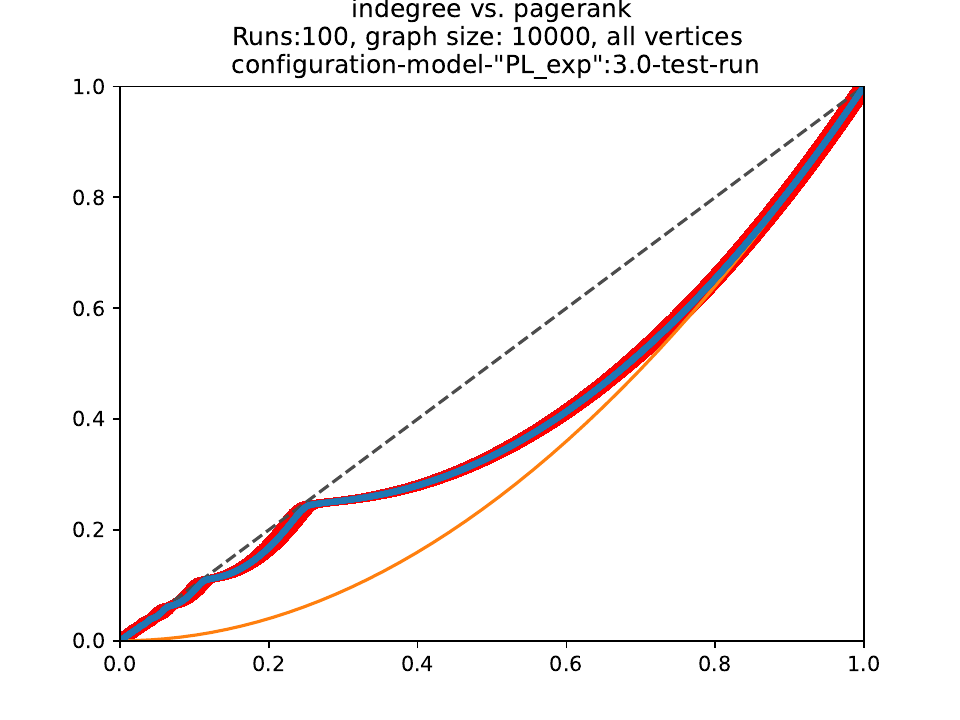}}
        \caption{Effect of alternative tie-breaking rules on the CCC: CCC of degree versus degree for an undirected configuration model with 500 vertices of degree~3, 400 of degree~4, and 300 of degree~5, with the  {\em random} tie-breaking rule, and PageRank versus indegree for a directed configuration model with an alternative tie-breaking rule. 
Compare with Figure~\ref{fig-PageRank-degree}.}
    \label{fig:both_images}
\end{figure}

\smallskip
A {\em key property} of our total order is that, by construction, no reordering occurs for $R = S$, and hence $\CCC(x) = x,$ the diagonal line corresponding to perfect agreement, which is a desirable property. 
This property is lost under alternative rules, as illustrated in Figure~\ref{fig:both_images}(a). In contrast, random tie-breaking can inflate apparent disagreement, pushing $\CCC(x)$ closer to $x^2$, which is the curve corresponding to near-independence.

Finally, other tie-breaking rules are arguably less natural. Indeed, 
purely random tie-breaking increases apparent disagreement when ties are common, while deterministic rules unrelated to the other measure may artificially reduce overlap without reflecting true structural differences. 
The hierarchical tie-breaking rule $(R,S,\mathrm{Uniform})$ thus provides a consistent, interpretable, and symmetric foundation for comparing centrality rankings via the CCC.

\subsection{General properties of CCC}
\label{sec-general-properties-CCC}
In this section, we provide the rationale behind several properties of the CCC curve discussed in Section~\ref{sec-ccc-maths}. 
First, the CCC is \emph{symmetric}, since its definition~\eqref{CCC-def} is symmetric in the two centrality measures. 
Second, the CCC curve always lies below the identity line, which follows directly from the definition. 
Furthermore, if one of the rankings is random, then $\CCC(x)$ represents the expected fraction of vertices appearing in the top $x\!\cdot\!100\%$ according to both rankings. 
For large~$n$, this fraction is approximately $x^2$. 
The same behaviour arises when one centrality measure corresponds to a random ordering of the vertices and the other induces a {\em deterministic} ranking, i.e., a centrality measure that does not have ties. 
This observation justifies calling two centrality measures $R$ and $S$ \emph{nearly independent} when $\CCC(x)$ is close to $x^2$.

\subsection{Sparse graphs and local convergence}
Here we discuss the relation between local convergence of sparse graph sequences and the convergence of the CCC. A {\em rooted} graph is a pair $(G,v)$ where $v\in V_G$. A graph sequence $(G_n)_{n\geq 1}$ is said to converge {\em locally weakly} if the expected proportion of vertices whose neighbourhoods look like a specific rooted graph converges in probability to some limit. This is equivalent to the statement that, when $o_n\in V_{G_n}$ is chosen uniformly at random, the random rooted graph $(G_n,o_n)$ converges in distribution to a (random) limiting rooted graph, which can be finite or infinite. Similarly, a graph sequence $(G_n)_{n\geq 1}$ is said to converge {\em locally in probability} if the proportion of vertices whose neighbourhoods look like a specific rooted graph converges in probability to some limit. See \cite[Chapter 2]{Hofs24} for more information.

In what follows in this and the next section, we will make the dependence of our centrality measures on the underlying graph explicit, and write $R^{\sss(G)}(v)$ for the centrality of $v\in V_G$ in the graph $G$. We call a centrality measure $R^{\sss(G)}\colon V_G\to [0,\infty)$ {\em local} when $R^{\sss(G)}(v)$ is close to the centrality measure $R^{\sss(B_r(v))}$ at $v$, computed on the $r$-neighbourhood $B_r(v)$ of $v$ in the graph $G$. For a local centrality measure, if $(G_n)_{n\geq 1}$ converges locally, also $R^{\sss(G_n)}(o_n)$ converges in distribution to the root-centrality $R^{\sss(G)}(o)$ of the limiting rooted graph $(G,o)$. This is relatively simple to verify for (in-, out-) degree centrality, but it also holds for PageRank \cite{GarHofLit20}, and, under some conditions on the largest eigenvalue of the graph's adjacency matrix, to Katz centrality \cite{Pand25}. The way to see this, is to show that the neighbourhood of the vertex $o_n$, {\em jointly with the centrality measure as marks associated to the vertices}, converges in distribution. When this is the case, under mild conditions, also the neighbourhood of the vertex $o_n$, jointly with the {\em pair} of local centrality measure as bivariate marks associated to the vertices will converge, and this implies the convergence of the CCC. Thus, the CCC converges for pairs of {\em jointly} local centrality measures, which is a rather nice property. 

Obviously, the above only applies to pairs of {\em local} centrality measures, and there are many {\em global} (i.e., non-local) centrality measures, for which the above reasoning does not apply. Important examples are centrality measures related to {\em graph distances}, such as closeness, betweenness, and load centrality. At the same time, it may be the case that, for appropriate graph sequences, such centrality measures can be well approximated by local quantities (see Figures \ref{fig-local-betweenness}-\ref{fig-local-betweenness-pat}). 
This contributes a significant challenge, that we leave for further research.

\subsection{Dense graphs and graphon convergence}
We next discuss convergence properties of the $\CCC$ for dense graphs, in which the number of edges grows proportionally to the square of the network size. For dense graphs, the natural convergence notion is {\em graphon convergence} \cite{Lova12,Zhao23}.
In a dense graph, each of the vertices $u$ has a type $x_u$. We assume that $(x_u)_{u\in V_G}\in [0,1]^n$ is such that its empirical measure converges to the Lebesgue measure. There is a directed edge between vertex $u$ to vertex $v$ with probability $W(x_u,x_v)$, independently. The function $W$ is called the {\em graphon}. We consider {\em undirected} settings, where $W$ is symmetric, as well as directed settings, where $W$ may not be symmetric and the edge between $u$ and $v$ is independent of the edge between $v$ and $u$.
\smallskip

Let $R=R^{\sss(G)}$ and $S=S^{\sss(G)}$ now be two centrality measures on the graph $G$ defined in terms of the graphon $W$ as above. {\em Convergence} of the CCC for converging dense graph sequences for the pair of centrality measures $R$ and $S$ would entail that ${\sf CCC}_{R,S}^{\sss(G)}(x)\rightarrow {\sf C}(x)$ for some limiting ${\sf C}$ that only depends on the graphon $W$. For concreteness, we here focus on (in- and out-)degree, closeness and betweenness centrality. As discussed above, closeness and betweenness centrality are {\em not} local for sparse settings, so this would indicate that convergence is more general for dense graphs. 

It is instructive to restrict to graphons $W$ that are {\em strictly positive}, in that there exists an $a>0$ such that $W(x,y)\geq a$ for all $x,y\in[0,1].$ Also assume that the edges in the undirected graph $(G_n)_{n\geq 1}$ are present {\em independently}. Then, distances in $G_n$ are rather well-behaved. Indeed, all distances in $G_n$ are 1 or 2 with high probability. Thus, there are $D_v$ vertices at distance 1 from $v\in V_G$, while there are $n-D_v-1$ vertices at distance 2 from $v\in V_G$. As such, the distances in the graph are closely related to the vertex degrees. Therefore, also the closeness centrality of $v$ is almost determined by $D_v$, and similarly for related distance-based centrality notions such as betweenness and load centrality. Since degrees are continuous in the graphon convergence topology, this should imply convergence of the CCC as well, as we clearly see in the simulations, and we discuss next. 

We simulated CCC graphs of growing sizes, wherein for each simulated size we computed the CCC for several graphs for several pairs of centrality measures, and with $G$ sampled from the random graph based on several undirected and directed graphons. To simulate dense graphs, we take $n$ vertices, fix their labeling to be $\{1,\ldots, n\}$, and, for every vertex, $v$, we draw an independent standard uniform $x_v$. Then, in general, we independently draw an edge between two vertices $i$ and $j$ with probability $W(x_i,x_j)$ for some $W,$ which depends on the setting, as in the sampling procedure in \cite{CaiAckFre16}.

\begin{figure}[h!]
\centering
\tempoff{\includegraphics[width=0.32\textwidth]{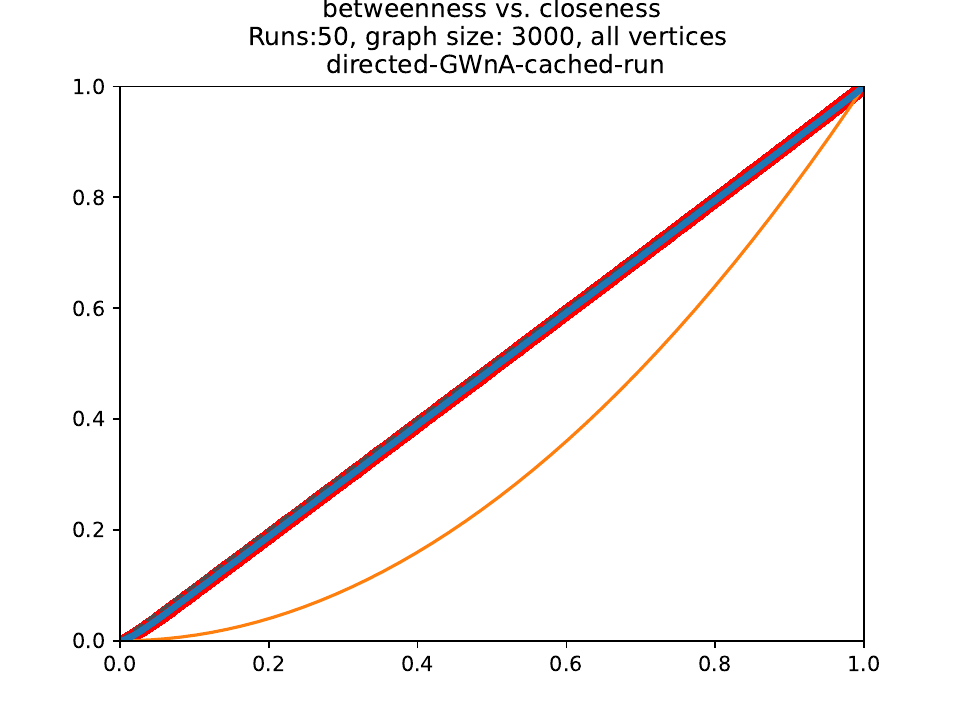}}
\tempoff{\includegraphics[width=0.322\textwidth]{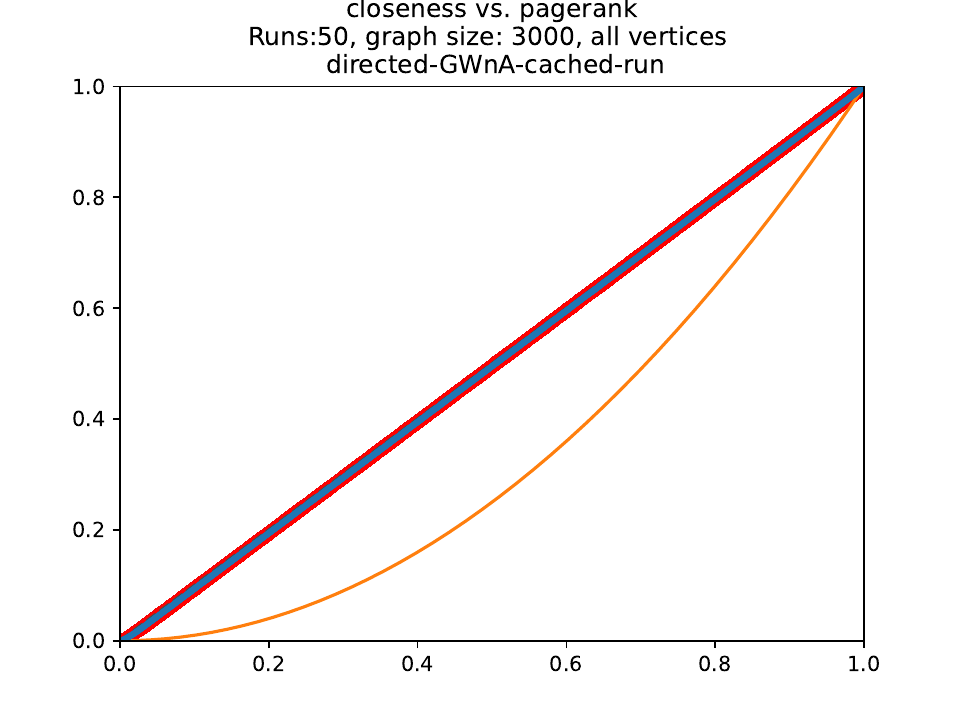}}
\tempoff{\includegraphics[width=0.32\textwidth]{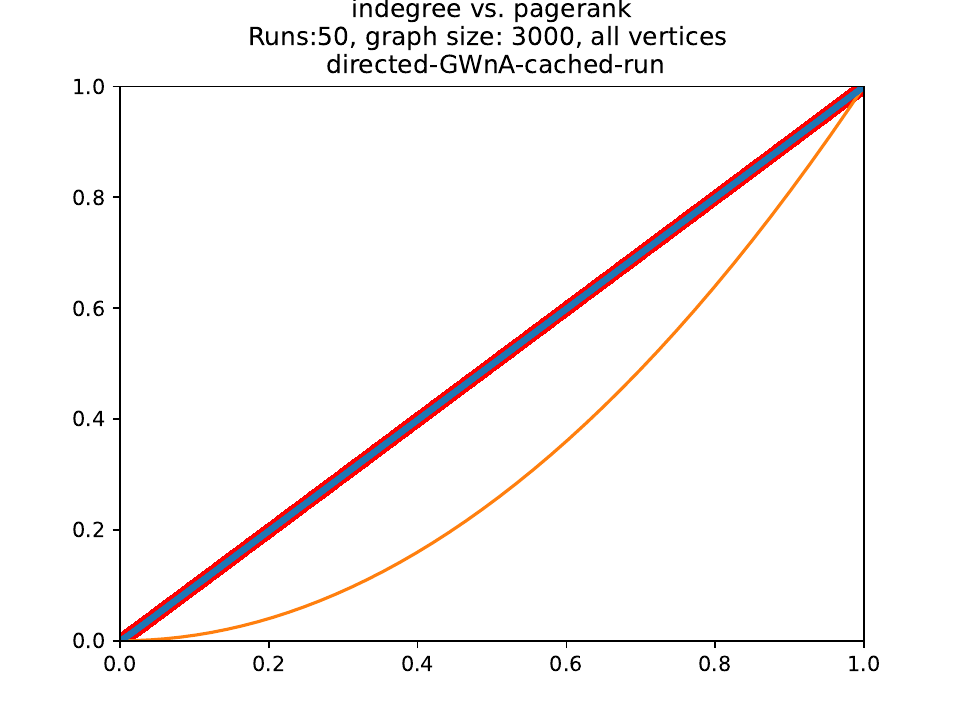}}
\caption{CCC for closeness versus betweenness, closeness versus in-degree, and in-degree versus PageRank for the undirected product graphon. All centralities are identical.}
\label{fig-undirected-dense-graph-product}
\end{figure}

For the undirected graphon, we use the product graphon $W(x,y) = cxy,$ for $c<1$. In Figure \ref{fig-undirected-dense-graph-product}, we see that all centralities are identical, which extends to {\em all} other pairs of centrality measures, figures of which can be found in \verb|https://doi.org/10.5281/zenodo.18214457|. There, also figures of the CCC for the sum graphon $W(x,y)=c(x+y)$ for $c<\tfrac{1}{2}$ can be found, which are again all the identity.

For the first directed graphon, independently for all $i<j$, a directed edge from vertex $i$ to $j$ is present with probability $W_{01}(x_i,x_j) = cx_ix_j, c<1$, and from vertex $j$ to $i$ $W_{10}(x_i,x_j) = c(1-x_i)(1-x_j)$, making in- and out-degrees maximally different. The results can be seen in Figure

\begin{figure}[h!]
\centering
\tempoff{\includegraphics[width=0.32\textwidth]{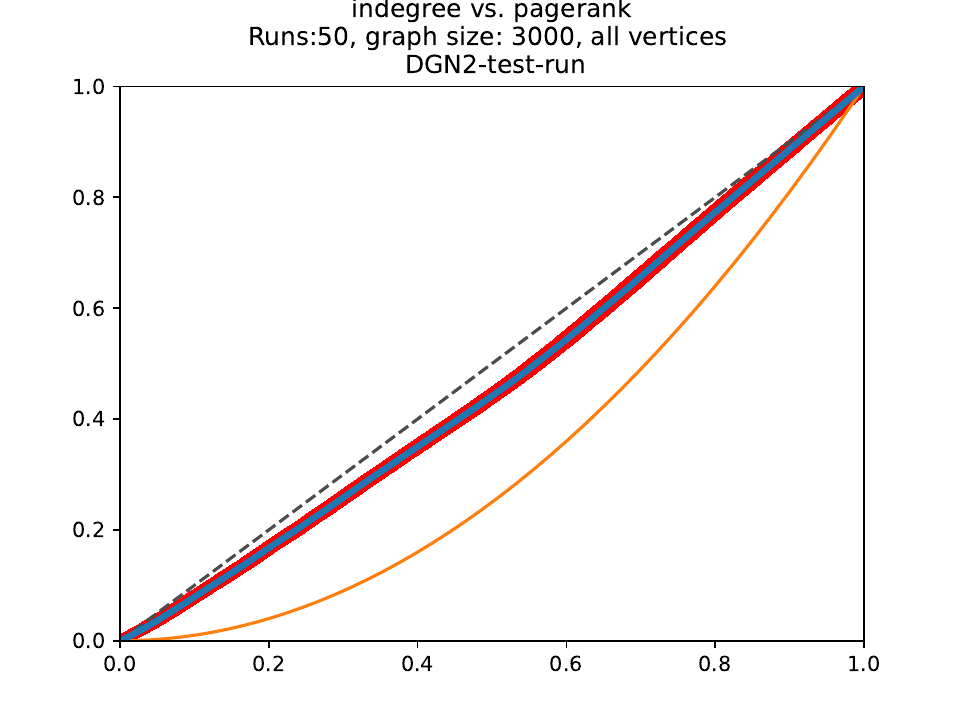}}
\tempoff{\includegraphics[width=0.32\textwidth]{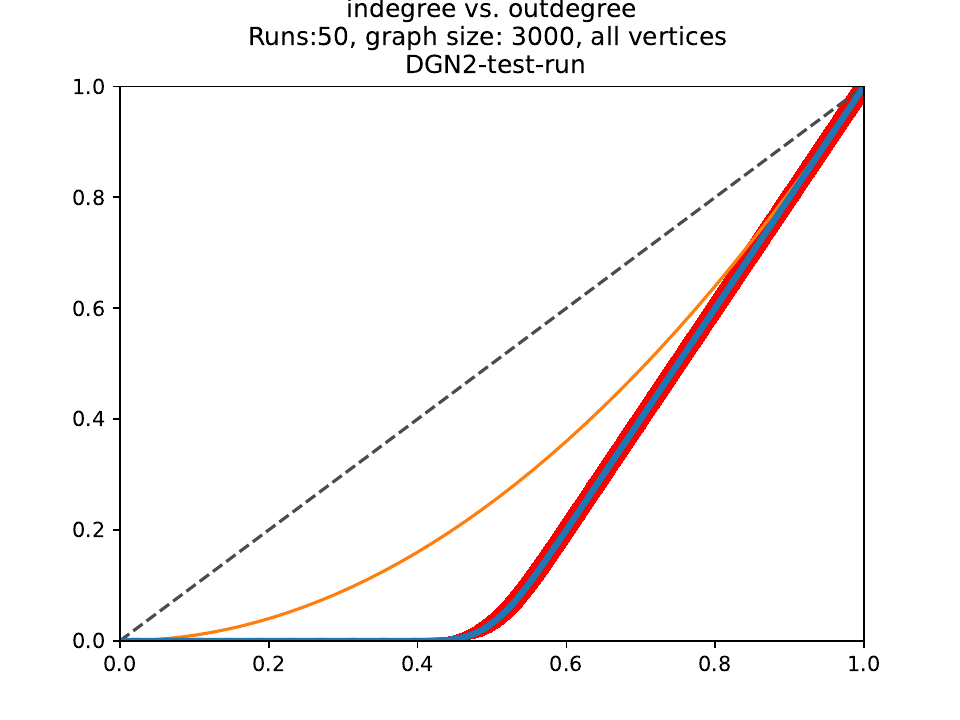}}
\tempoff{\includegraphics[width=0.32\textwidth]{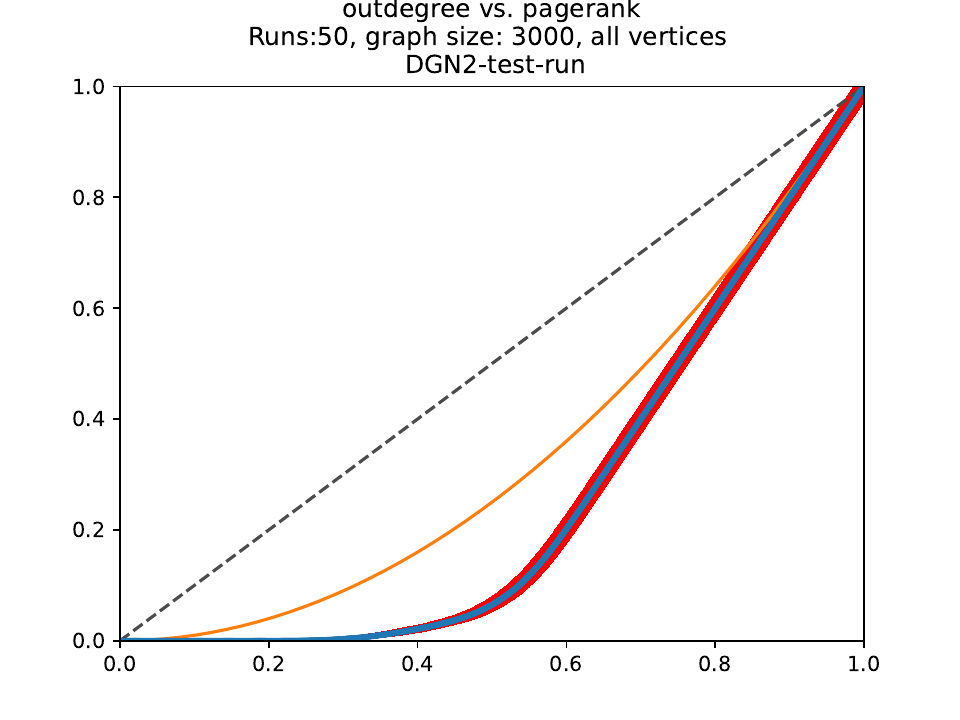}}
\caption{CCC for in-degree versus out-degree, in-degree versus PageRank, and out-degree versus PageRank for the directed product graphon where in-degrees and out-degrees are maximally different. In-degree and PageRank are almost identical, out-degree is close to maximally different from in-degree and PageRank.}
\label{fig-undirected-dense-graph-different}
\end{figure}

\begin{figure}[h!]
\centering
\tempoff{\includegraphics[width=0.32\textwidth]{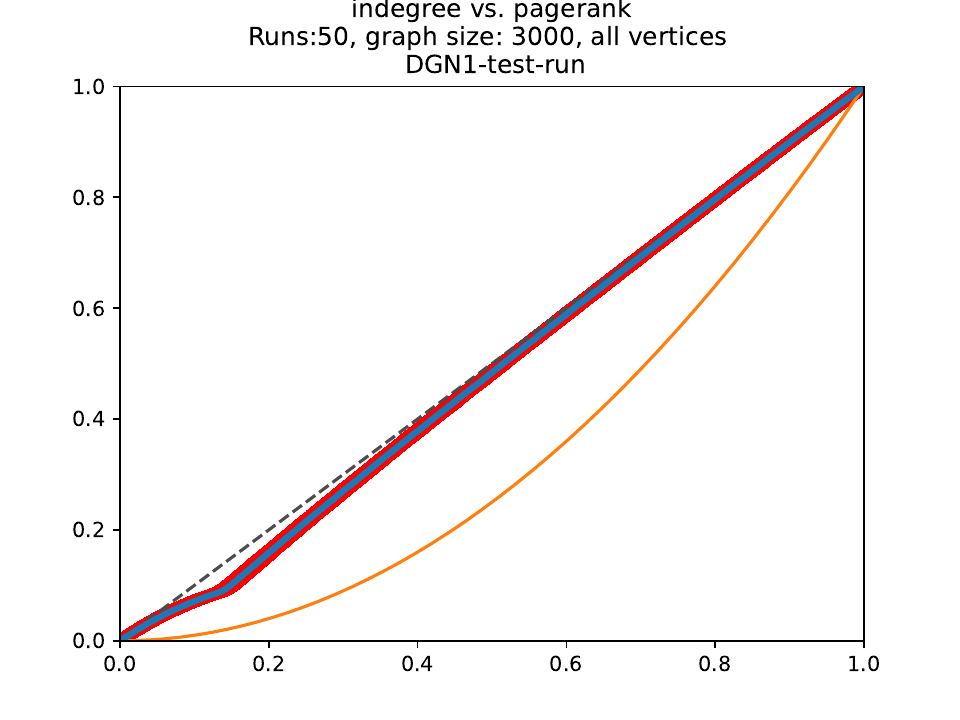}}
\tempoff{\includegraphics[width=0.32\textwidth]{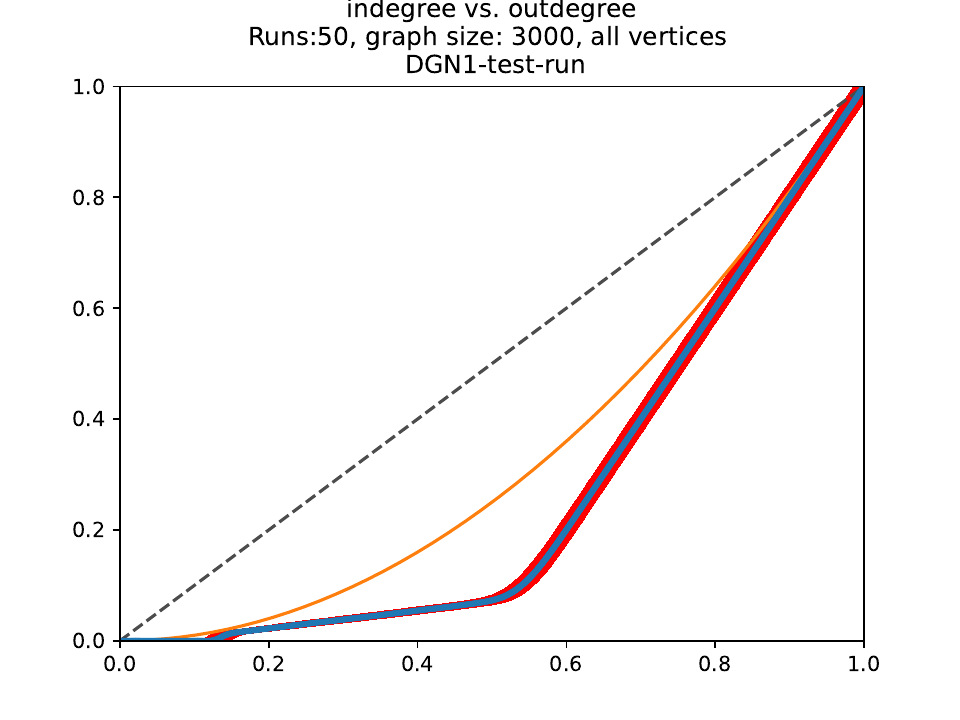}}
\tempoff{\includegraphics[width=0.32\textwidth]{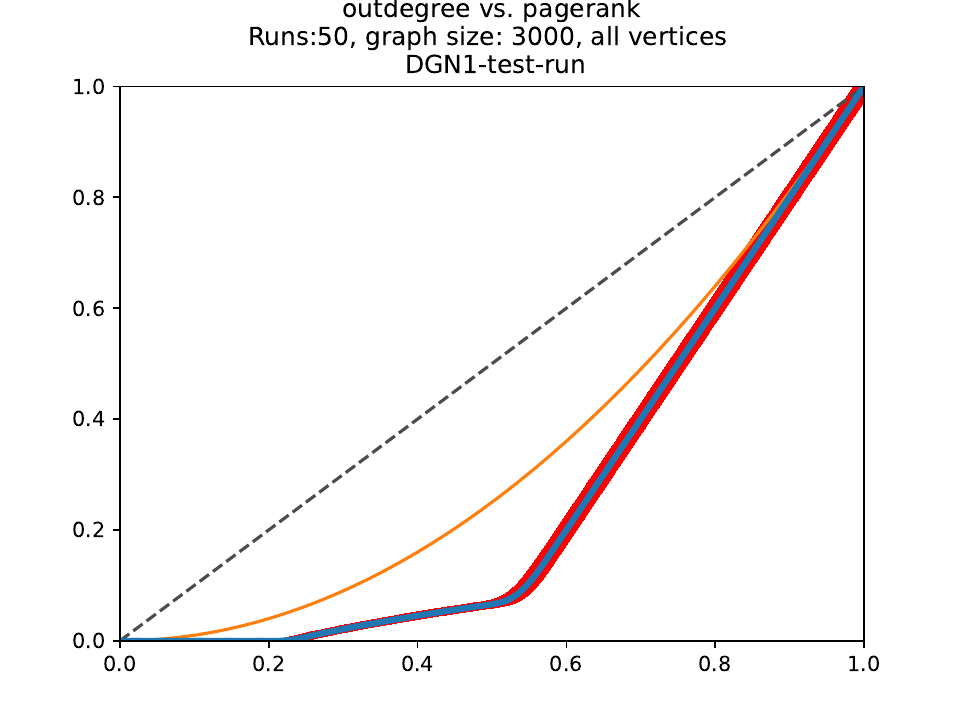}}
\caption{CCC for in-degree versus PageRank, in-degree versus out-degree, and out-degree versus PageRank for the \cite{CaiAckFre16}.}
\label{fig-dense-graph-CaiAckFre16}
\end{figure}

For the directed graphons proposed in \cite{CaiAckFre16}, 
independently, the directed edge from $i$ to $j$ is present with probability $W_{10}(x_i, x_j) = c_{\sss \rm high}$ if $x_j < p$, and $c_{\sss \rm low}$ otherwise, where the constants $c_{\sss \rm high},c_{\sss \rm low}$ and $p$ are chosen as $c_{\sss \rm high}=0.9$, $c_{\sss \rm low}=0.05$, and $p=0.15$. Figure \ref{fig-dense-graph-CaiAckFre16} shows that PageRank is similar to in-degree, while PageRank and in-degree are both almost maximally different from out-degree.

We conclude that the amount of variability in the CCCs in different samples is minute, thus suggesting that convergence indeed holds for most pairs of centrality measures. Also, for undirected graphs, we see that the CCC is often either close to the identity curve, or to the maximally different curve, indicating that many centrality measures are similar, or rather maximally different, for dense graphs, which would be an important conclusion.

\section*{\bf Acknowledgments.}
The work in this paper was supported by the Netherlands Organisation for Scientific Research (NWO) through Gravitation-grant NETWORKS-024.002.003. The work of MP was carried out in part during his previous employment at Eindhoven University of Technology. The work of ON was carried out in part during his previous employment at Leiden University. 

\bibliography{refs}
\bibliographystyle{abbrv}

\newpage
\appendix

\section{Supplementary material}
\subsection{Definitions of popular centrality measures}
Centrality measures in networks are numbers assigned to vertices indicating their importance.
We now formally define what a centrality measure is. Let $G=(V_G, E_G)$ be a graph/digraph. A centrality measure $R\colon V_G\to \mathbb{R}_{\geq0}$, is a function from the vertex set to positive real numbers, where the value at a vertex indicates how important it is. 

There are many  centrality measures that have been considered in the literature. In this paper, we consider in-degree, out-degree, total degree, PageRank, betweenness centrality, closeness centrality, load centrality, harmonic centrality, Katz centrality, eigenvector centrality; both in directed and undirected graphs. 
\smallskip

\paragraph*{\bf PageRank.} Let $R\colon V_G \to \mathbb{R}_\geq0$ be the PageRank function, which satisfies the recursion 
    \eqn{
    \label{PageRank-def} R(i) = c\sum_{j \in V_G}\frac{e_{j,i}}{d_j^+}R(j) + (1-c),
    }
where $e_{j, i}$ is the number of edges from $j$ to $i$ in $G$, $d_j^+$ is the out-degree of $j$, and $c \in (0, 1)$ is the damping factor.
\smallskip

\paragraph*{\bf Katz centrality.} Katz centrality $K_{\alpha}: V_G \to \mathbb{R}_\geq0$, introduced by Leo Katz \cite{katz1953new}, is a centrality measure defined as 
    \eqn{
    \label{Katz-def} K_\alpha(i) = \sum_{k \geq 1}\sum_{j \in V_G} \alpha^{k}A^k_{i,j},
    }
where $A$ is the adjacency matrix of $G$ and $\alpha \in (0, 1)$ is the attenuation factor.

Notice that in order for the sum in \eqref{Katz-def} to converge, the value of the attenuation factor $\alpha$ has to be chosen such that it is smaller than the reciprocal of the largest eigenvalue of $A$ (which is non-negative, due to the non-negativity of $A$).
\smallskip

\paragraph*{\bf Eigenvector centrality.}
    Let $x\colon V_G \to \mathbb{R}_\geq0$ be the eigenvector centrality vector, then it satisfies
    \eqn{
    \label{eigenvector-def}
    x^t A = \lambda x^t,
    }
    where $A$ is the adjacency matrix of $G$ and $\lambda$ is its associated largest modulus eigenvalue. This eigenvector is unique up to a multiplicative factor. To define a unique score, one must normalise the eigenvector $x$. 
\smallskip

\paragraph*{\bf Betweenness centrality.} Let $b\colon V_G \to \mathbb{R}_\geq0$ be the betweenness centrality function, then
    \eqn{
    \label{betweenness-def}
    b(v) = \sum_{i \in V_G}\sum_{j \in V_G}\frac{\sigma_{i,j}(v)}{\sigma_{i, j}},
    }
where $\sigma_{i, j}(v)$ counts the number of shortest paths from $i$ to $j$ containing $v$ and $\sigma_{i, j}$ is the number of shortest paths from $i$ to $j$. For betweenness$k$, we restrict in the above to pairs of vertices $i,j\in V_G$ such that $i$ and $j$ are at graph distance at most $k$. 
\smallskip

\paragraph*{\bf Load centrality.} Let $l\colon V_G \to \mathbb{R}_\geq0$, the load centrality function, then 
     \eqn{
     \label{load-def}
     l(v) = \frac{\sum_{i \in V_G}\sum_{j \in V_G}\sigma_{i,j}(v)}{\sum_{i \in V_G}\sum_{j \in V_G}\sigma_{i, j}},
     }
where $\sigma_{i, j}(v)$ counts the number of shortest paths from $i$ to $j$ containing $v$ and $\sigma_{i, j}$ is the number of shortest paths from $i$ to $j$. For load$k$, we restrict in the above to pairs of vertices $i,j\in V_G$ such that $i$ and $j$ are at graph distance at most $k$. 
\smallskip

\paragraph*{\bf Closeness centrality.} Let $c \colon V_G \to \mathbb{R}_\geq0$ be the closeness centrality function, then 
    \eqn{
    \label{closeness-def}
    c(v) = \frac{|V_G|-1}{\sum_{u \in V_G}d_{G}(u, v)},
    }
where $d_{G}(u, v)$ is the length of the shortest path between $u$ and $v$ in $G$.
\smallskip

\paragraph*{\bf Harmonic centrality.} 
    Let $h\colon V_G \to \mathbb{R}_{\geq0}$ be the harmonic centrality function, then
    \eqn{
    \label{harmonic-def}
    h(v) = \sum_{ u\neq v}\frac{1}{d_G(u, v)},
    }
where $d_{G}(u, v)$ is the length of the shortest path between $u$ and $v$ in $G$.

\subsection{Details on the artificial networks used}
The source code used to create plots in the main article is available on GitHub, via \verb|https://github.com/nagyol/CCC|. The program makes extensive use of the NetworkX and igraph libraries \cite{networkx}.

\paragraph{\bf Simulating directed configuration models.} In the case of (un-)directed configuration model with power-law degree sequence with exponent $\alpha$, the individual degrees are chosen to be
\begin{align*}
    \deg(v) = \lfloor d_{\rm min} + Y \rfloor,
\end{align*}
where $d_{\rm min}$, the lower-bound on degrees, is chosen to be 1, and $Y$ is a $\mathsf{Pareto}(\alpha-1)$-distributed random variable, i.e., random variable with CDF
\begin{align*}
    F_Y(x) = \begin{cases}
        1 - \frac{1}{x^\alpha} & x \geq 1,\\
        0 & \text{otherwise}.
    \end{cases}
\end{align*}

In the case of the {\em directed} configuration model, first the in-degree sequence is sampled via the procedure outlined above, and then the out-degree sequence is resampled until it has the same sum as the in-degree sequence. 
\smallskip

\paragraph{\bf Simulating directed graphons.} In the case of graphon-based simulations, the focus was on illustrating the conjectured limit of CCC as the size of the individual graphs tends to infinity. The individual graphs were sampled according to the following procedure:
\begin{enumerate}
\item Create an empty graph with $n$ vertices at the beginning. 
\item For each vertex $v\in V_G$, assign a weight $x_v \sim \mathsf{Unif}([0,1])$.
\item Create an edge between edges $u,v\in V_G$ based on an outcome of an independent $\mathsf{Bernoulli}(cx_u x_v)$ trial. 
\end{enumerate}
This random graph is also known as the $G(W,n)$ model (see e.g., \cite{Lova12, Zhao23}).

\subsection{CCC for undirected graphs}
In this section, we present the same figures as in the main text of the paper, but now for {\em undirected} real-world and artificial  networks. Also in such networks, centrality measures are highly used and useful to identify important vertices in the network. The real-world network consists of the collaboration network studied in \cite{LKF2007}, while the artificial networks correspond to configuration models with power-law exponent $3$.
\begin{figure}[H]
\centering
\tempoff{\includegraphics[width=0.5\columnwidth]{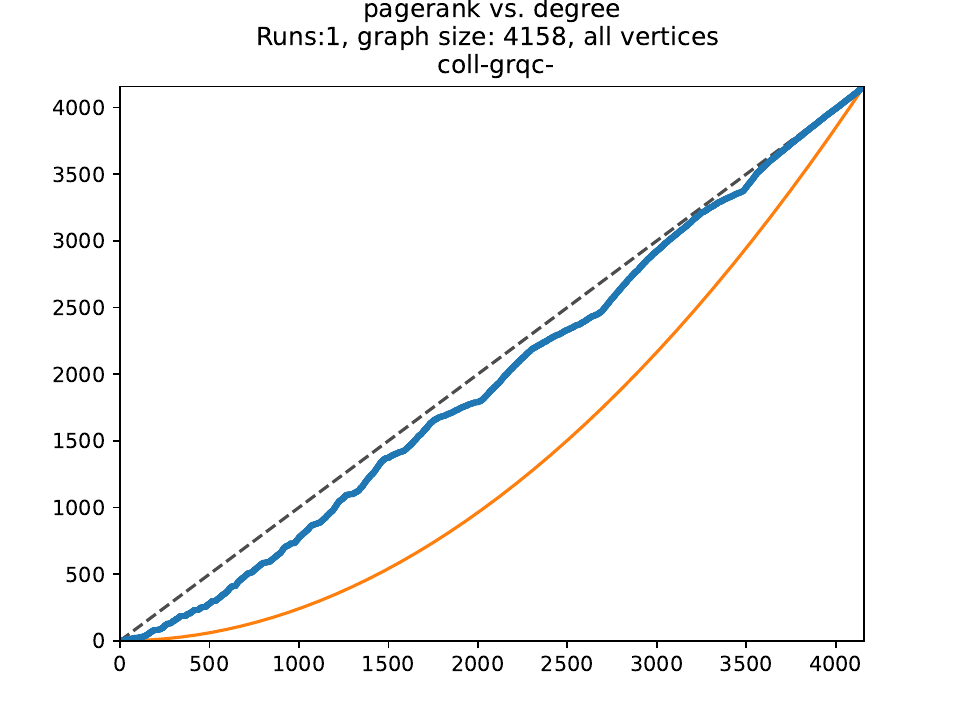}}
    \caption{CCC for PageRank versus degree for the collaboration network \cite{LKF2007}.}
    \label{fig-undir-collaboration-network-PageRank-Degree}
\end{figure}
\begin{figure}[H]
\centering
\tempoff{\includegraphics[width=0.45\columnwidth]{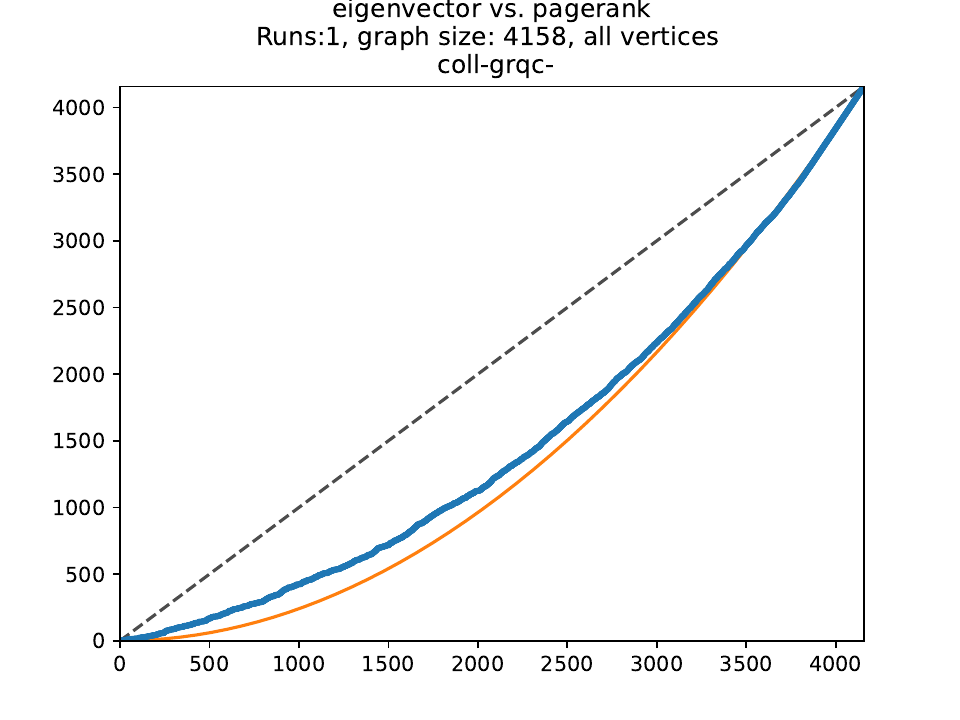}}
\tempoff{\includegraphics[width=0.45\columnwidth]{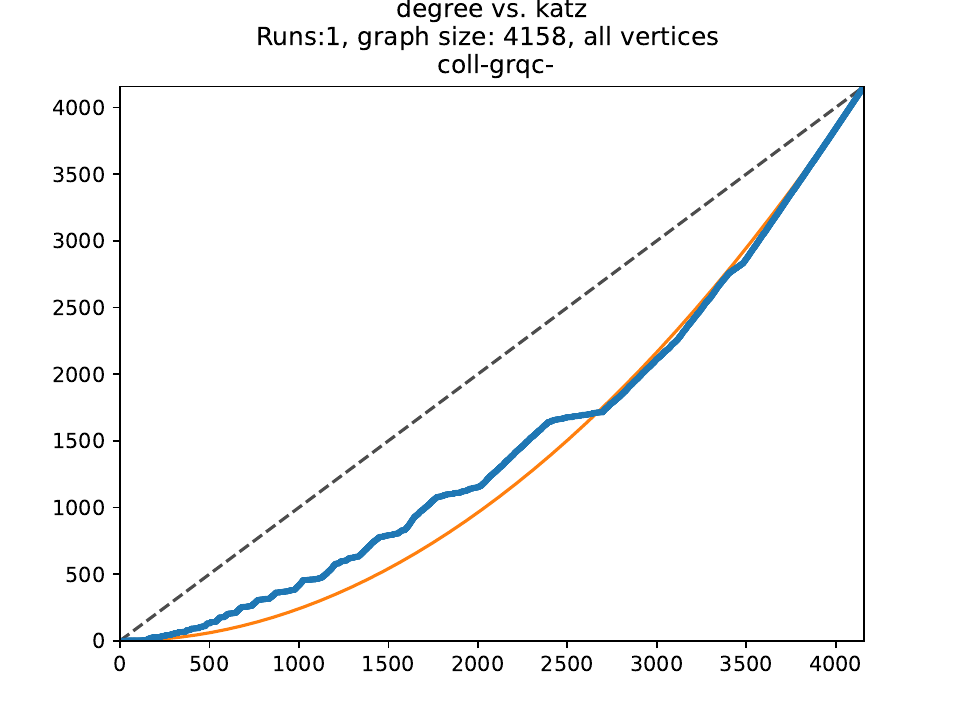}}
    \caption{CCC for eigenvector versus PageRank, and degree versus Katz centrality, respectively, for the collaboration network \cite{LKF2007}.}
    \label{fig-undir-collaboration-network-others}
\end{figure}
\FloatBarrier
\begin{figure}[H]
\centering
\tempoff{\includegraphics[width=0.45\columnwidth]{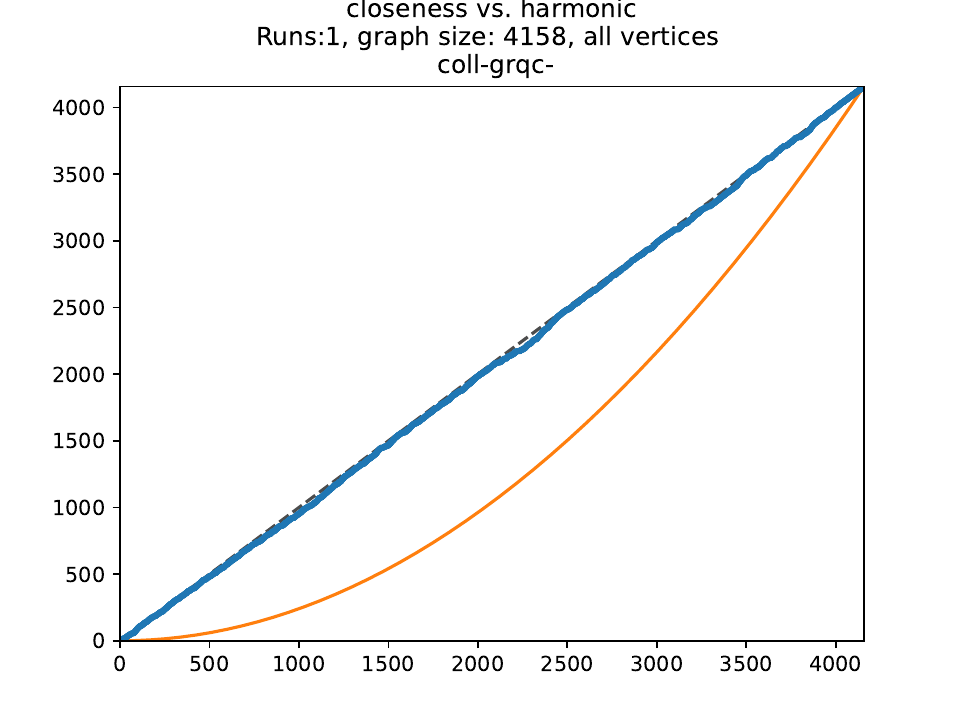}}
\tempoff{\includegraphics[width=0.45\columnwidth]{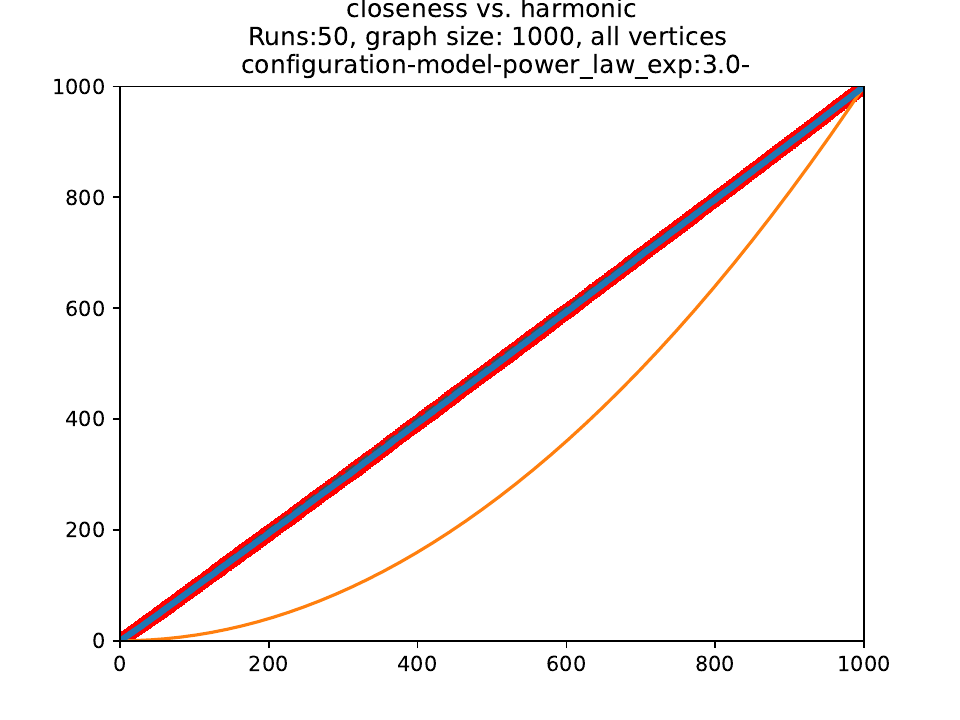}}
    \caption{CCC for closeness versus harmonic centrality for the collaboration network \cite{LKF2007} and the artificial network.}
    \label{fig-undir-closeness-harmonic}
\end{figure}
\FloatBarrier
\begin{figure}[H]
\centering
\tempoff{\includegraphics[width=0.45\columnwidth]{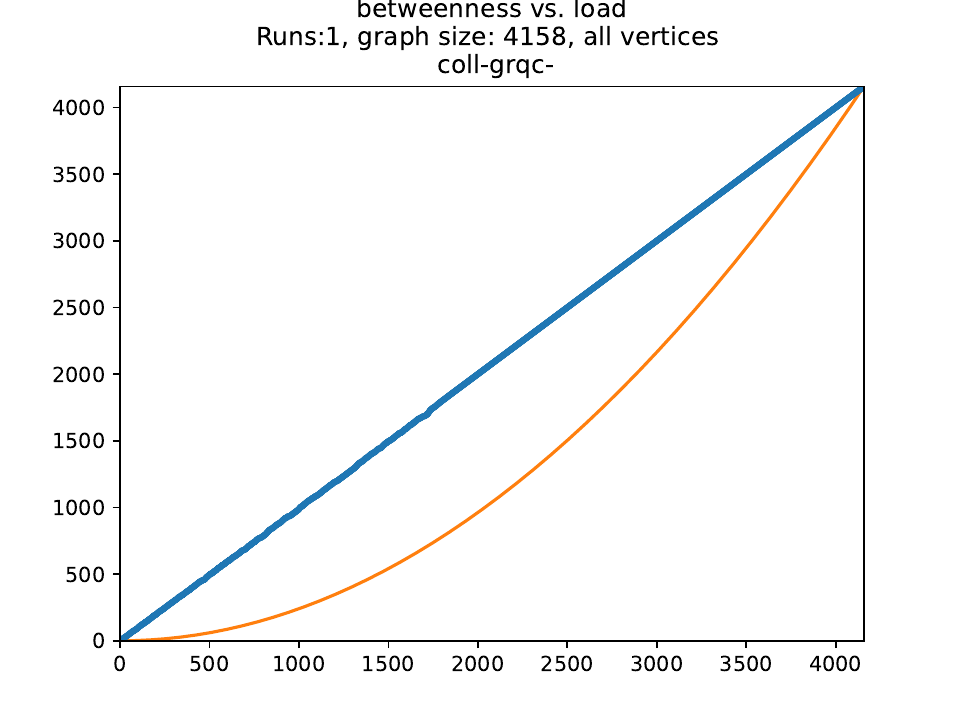}}
\tempoff{\includegraphics[width=0.45\columnwidth]{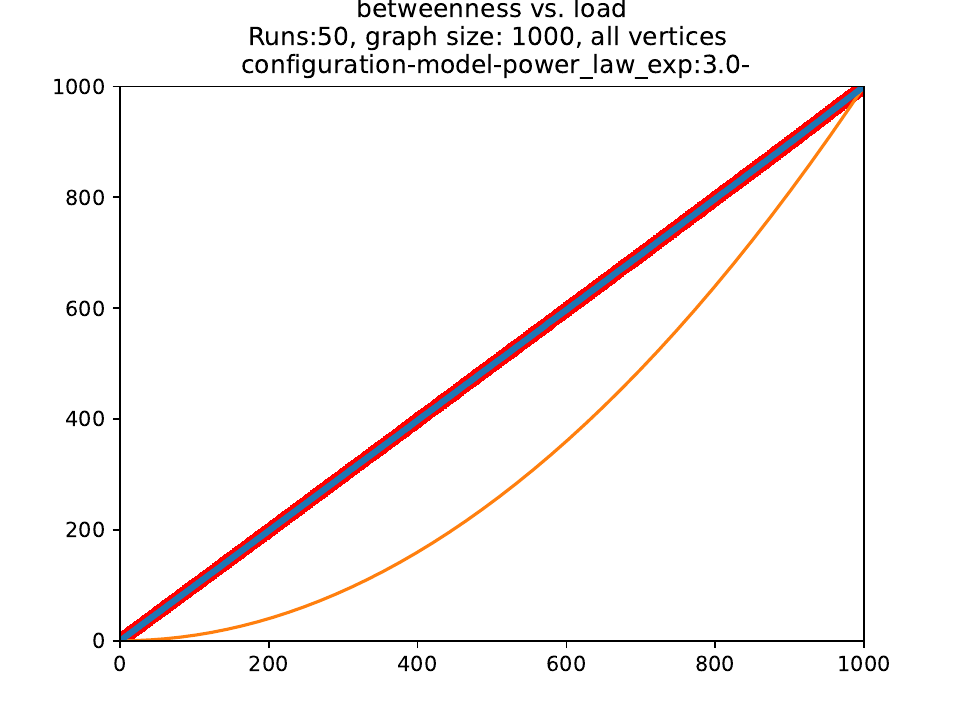}}
    \caption{CCC for betweenness versus load centrality for the collaboration network \cite{LKF2007} and the artificial network.}
    \label{fig-undir-betweenness-load}
\end{figure}
\FloatBarrier
\begin{figure}[H]
\centering
\tempoff{\includegraphics[width=0.45\columnwidth]{figures/pagerank-degree-coll-grcq.pdf}}
\tempoff{\includegraphics[width=0.45\columnwidth]{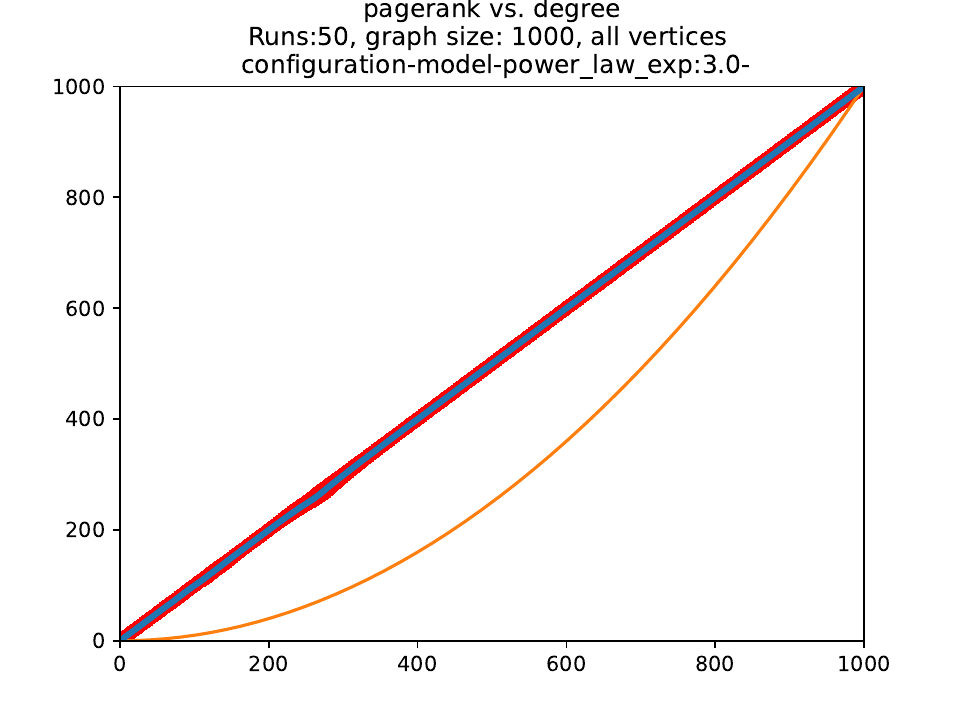}}
    \caption{CCC for PageRank versus degree for the collaboration network \cite{LKF2007} and the artificial network.}
    \label{fig-undir-PageRank-degree}
\end{figure}
\FloatBarrier
\begin{figure}[H]
\centering
\tempoff{\includegraphics[width=0.45\columnwidth]{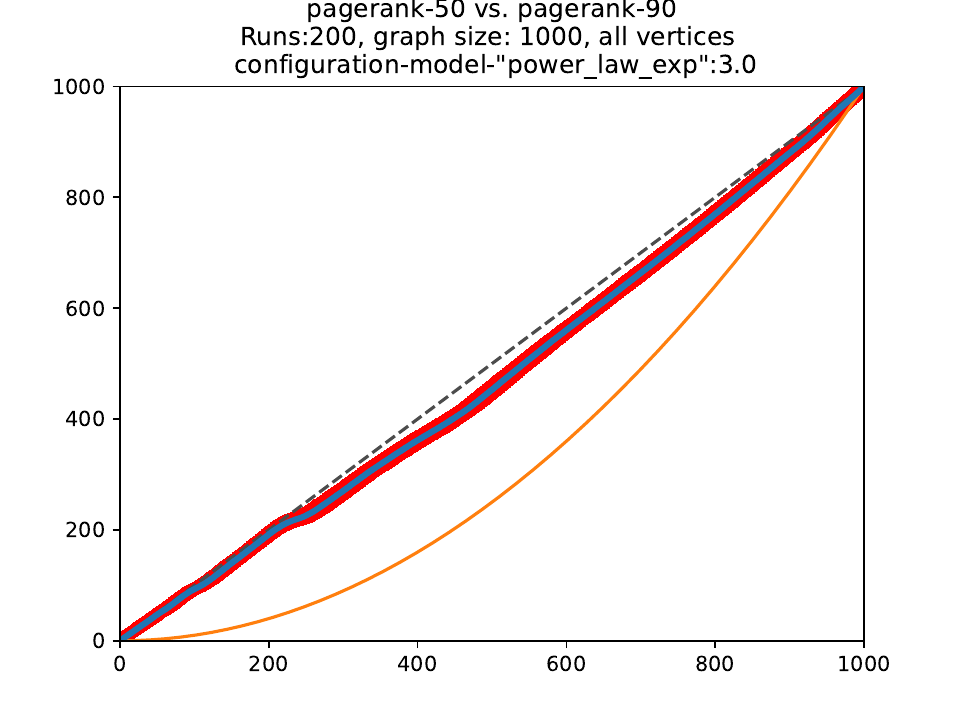}}
\tempoff{\includegraphics[width=0.45\columnwidth]{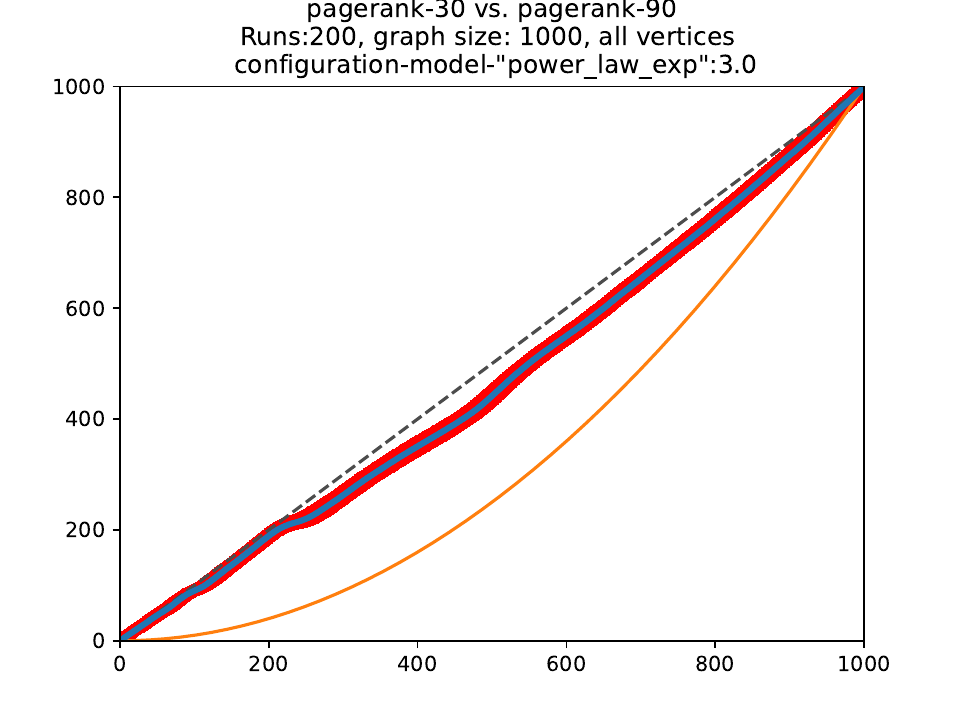}}
    \caption{CCC for PageRank with damping factors 0.5 versus 0.9, and 0.3 versus 0.9, respectively, for the artificial network.}
    \label{fig-undir-PageRank-damping}
\end{figure}
\begin{figure}[H]
\centering
\tempoff{\includegraphics[width=0.45\columnwidth]{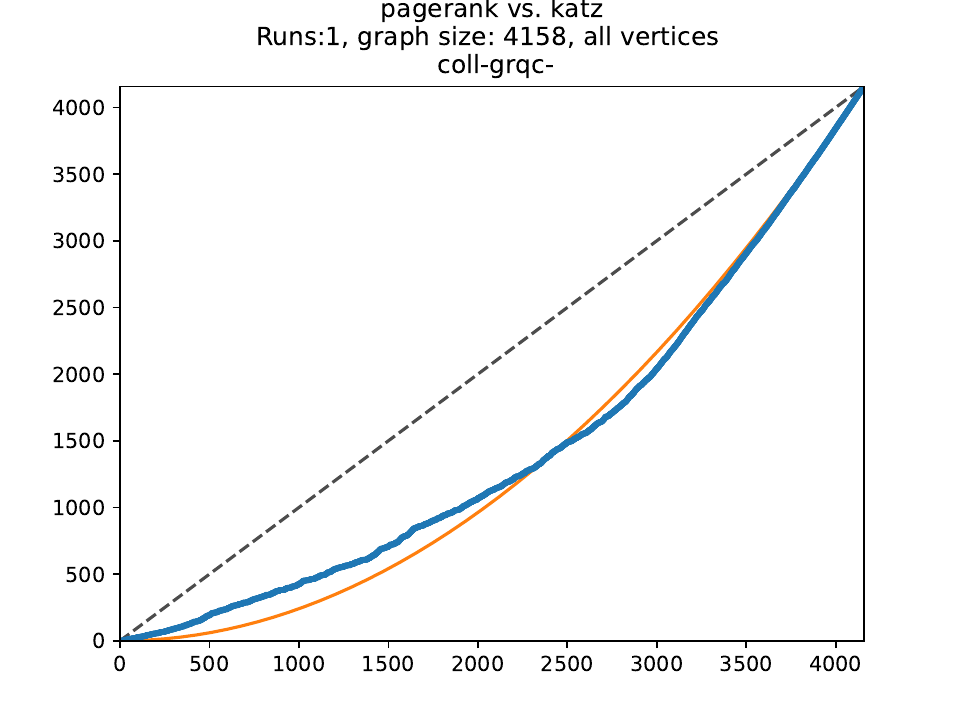}}
\tempoff{\includegraphics[width=0.45\columnwidth]{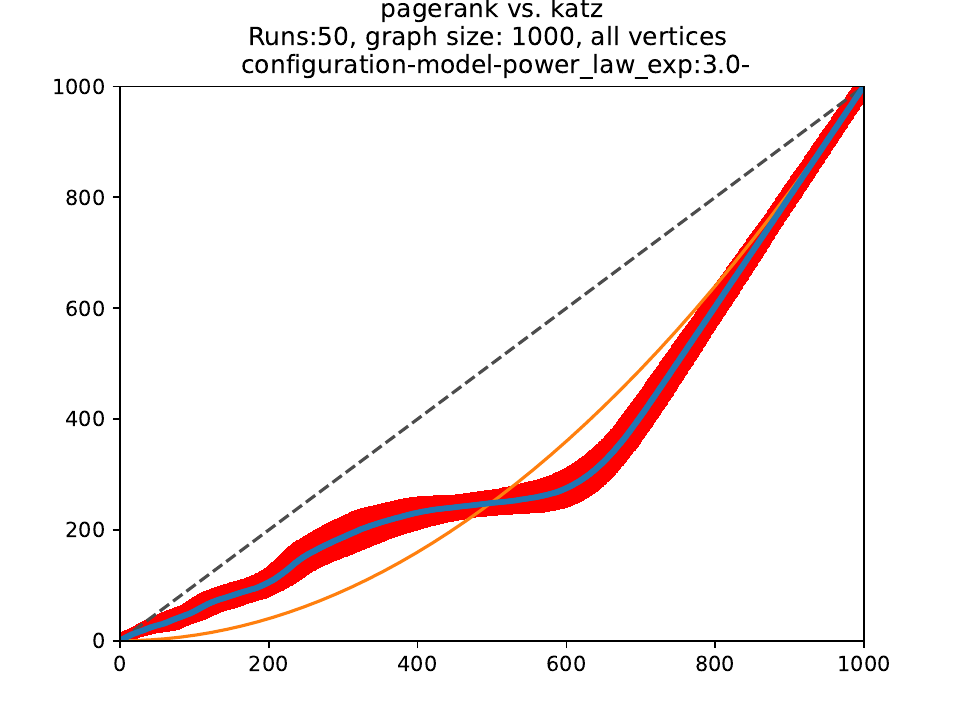}}
    \caption{CCC for PageRank versus Katz, for the collaboration network \cite{LKF2007} and the artificial network.}
    \label{fig-undir-PageRank-Katz}
\end{figure}

\subsection{The CCC for related directed real-world networks}\label{sec:appendix-directed-realworld-related}

In this subsection we present additional CCC plots for two large directed real-world networks that are structurally different from the citation network considered in the main text, namely the Stanford web graph and the U.S.\ patent citation network. The goal is to illustrate that the CCC can reveal both strong alignment and near-independence between centrality measures, depending on the dataset and the pair of measures.

\smallskip
\paragraph{\bf Stanford web graph.}
For the Stanford dataset, the CCC for betweenness versus closeness lies close to the independence curve $x^{2}$, indicating that these two measures identify largely disjoint
sets of highly ranked vertices. A similar near-independence is observed for the pairs betweenness versus harmonic, closeness versus PageRank, and harmonic versus PageRank.

\vspace{-0.2cm}
\begin{figure}[H]
\centering
\tempoff{\includegraphics[width=0.45\textwidth]{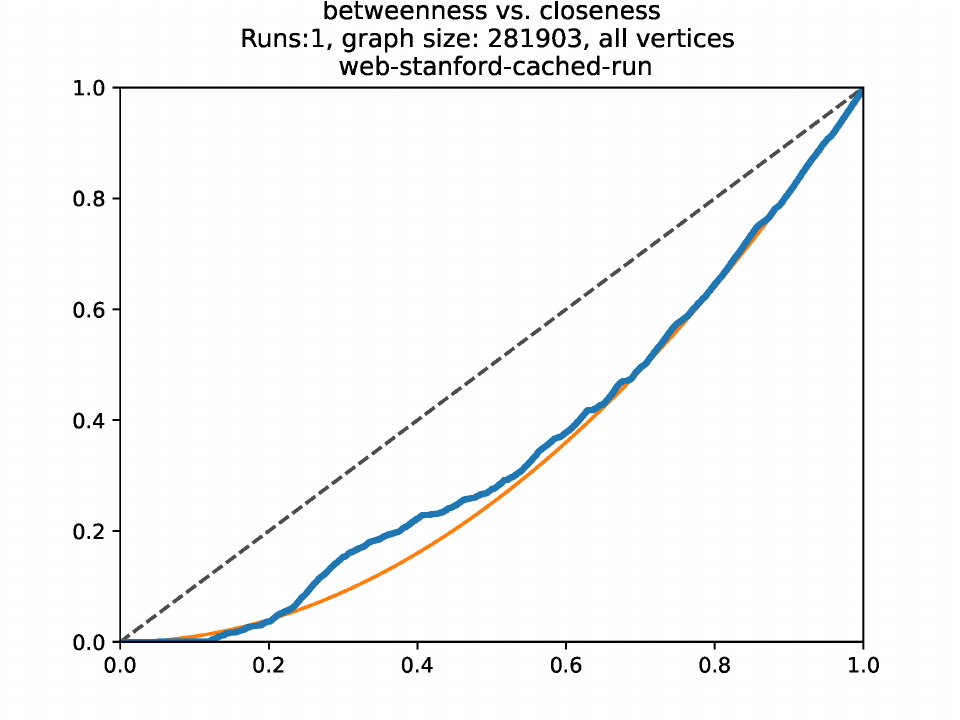}}
\tempoff{\includegraphics[width=0.45\textwidth]{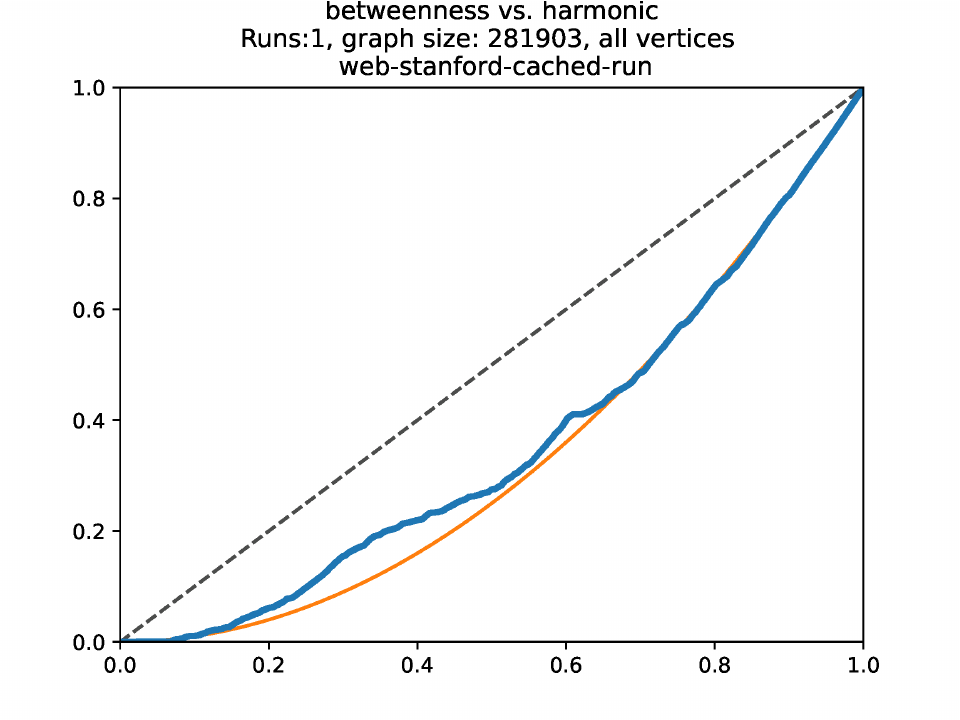}}
\caption{Stanford web graph: CCC for betweenness versus closeness (left) and betweenness versus harmonic (right). Both curves are close to $x^2$, indicating near-independence.}
\label{fig:stan-btw-close-harm}
\vspace{-0.2cm}
\end{figure}

\vspace{-0.2cm}
\begin{figure}[H]
\centering
\tempoff{\includegraphics[width=0.45\textwidth]{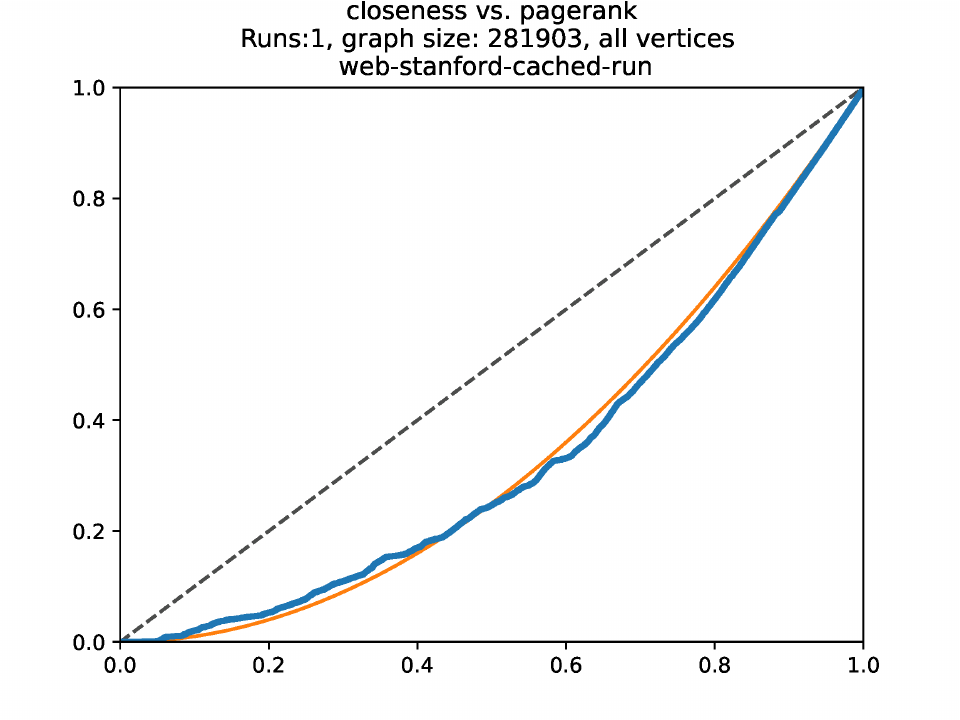}}
\tempoff{\includegraphics[width=0.45\textwidth]{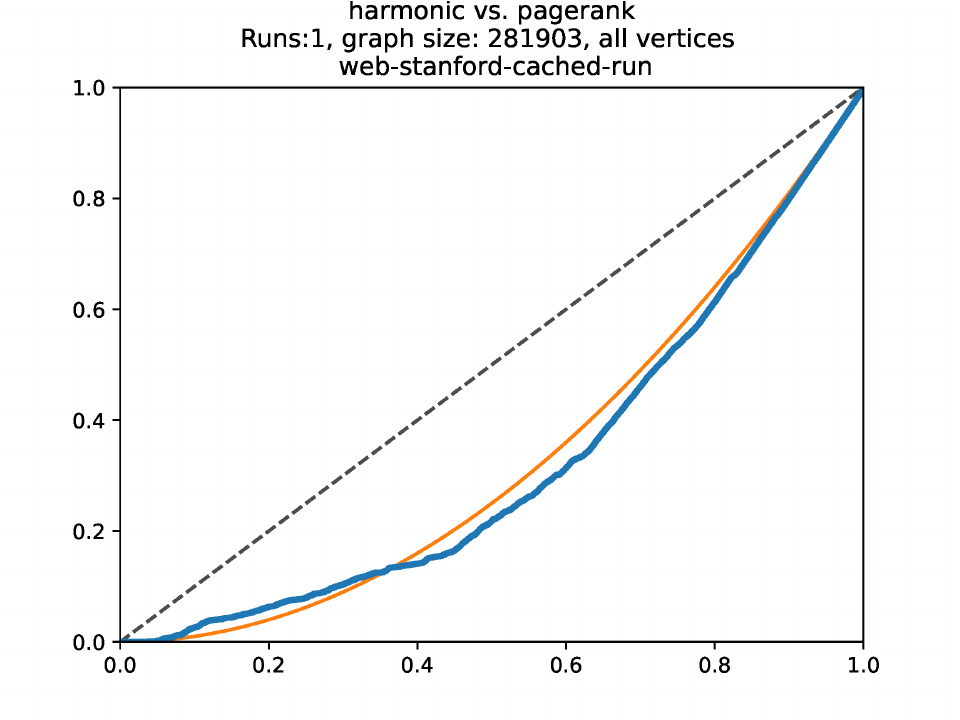}}
\caption{Stanford web graph: CCC for closeness versus PageRank (left) and harmonic versus PageRank (right). Both curves are close to $x^2$, indicating near-independence.}
\label{fig:stan-pr-close-harm}
\vspace{-0.2cm}
\end{figure}

\smallskip
\paragraph{\bf U.S.\ patent citation network.}
For the U.S.\ patent citation network, several pairs of directed centrality measures exhibit
near-independence. In particular, the CCCs for in-degree versus out-degree, out-degree versus
PageRank, and Katz versus out-degree are all close to $x^2$. This indicates that these pairs
identify substantially different sets of top-ranked vertices, highlighting that, in this
dataset, incoming- and outgoing-link structures encode rather distinct notions of importance.

In contrast, PageRank and in-degree display a markedly different behavior. Their CCC is
significantly above $x^2$, indicating strong dependence.

\vspace{-0.2cm}
\begin{figure}[H]
\centering
\tempoff{\includegraphics[width=0.45\textwidth]{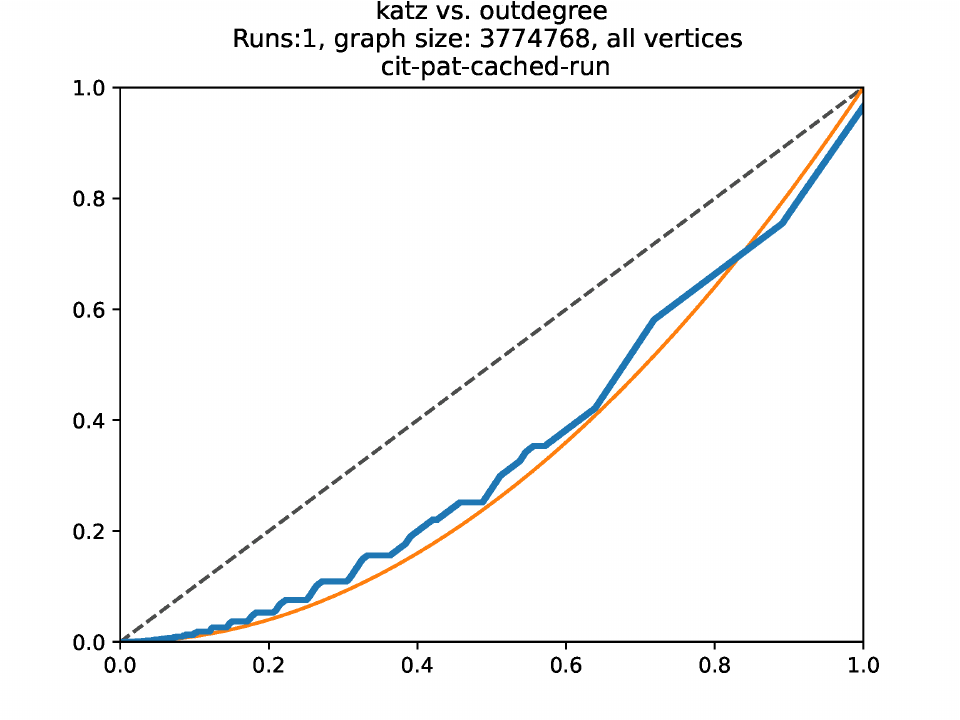}}
\tempoff{\includegraphics[width=0.45\textwidth]{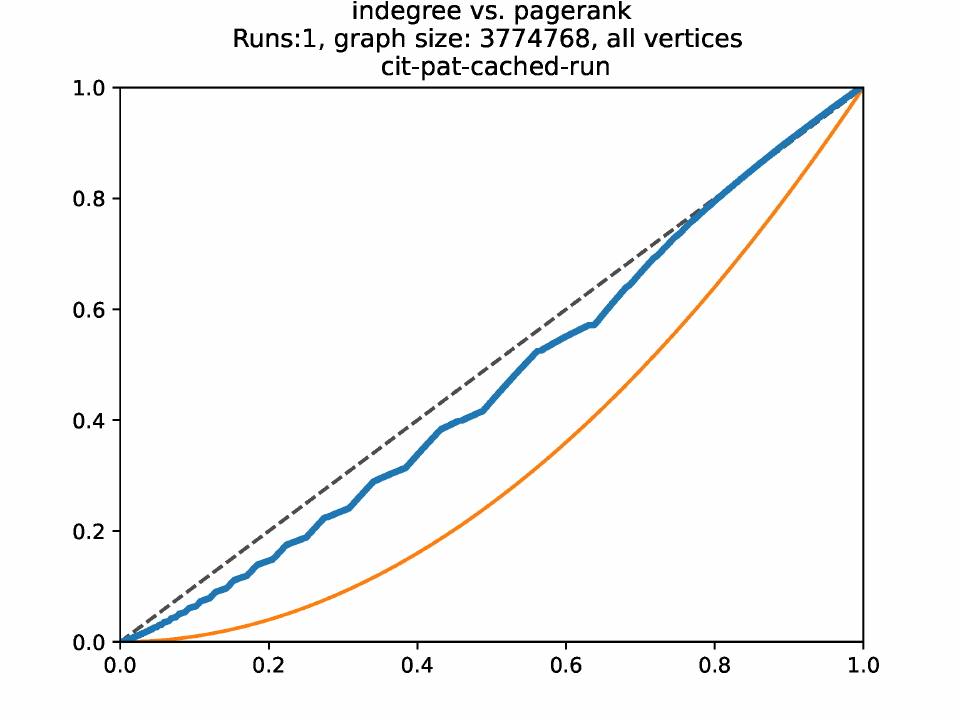}}
\caption{Patent network: CCC for Katz versus out-degree (left) and PageRank versus in-degree (right).
Katz and out-degree are nearly independent, while PageRank and in-degree exhibit strong dependence.}
\label{fig:pat-katz-out-pr-in}
\vspace{-0.2cm}
\end{figure}

\vspace{-0.2cm}
\begin{figure}[H]
\centering
\tempoff{\includegraphics[width=0.45\textwidth]{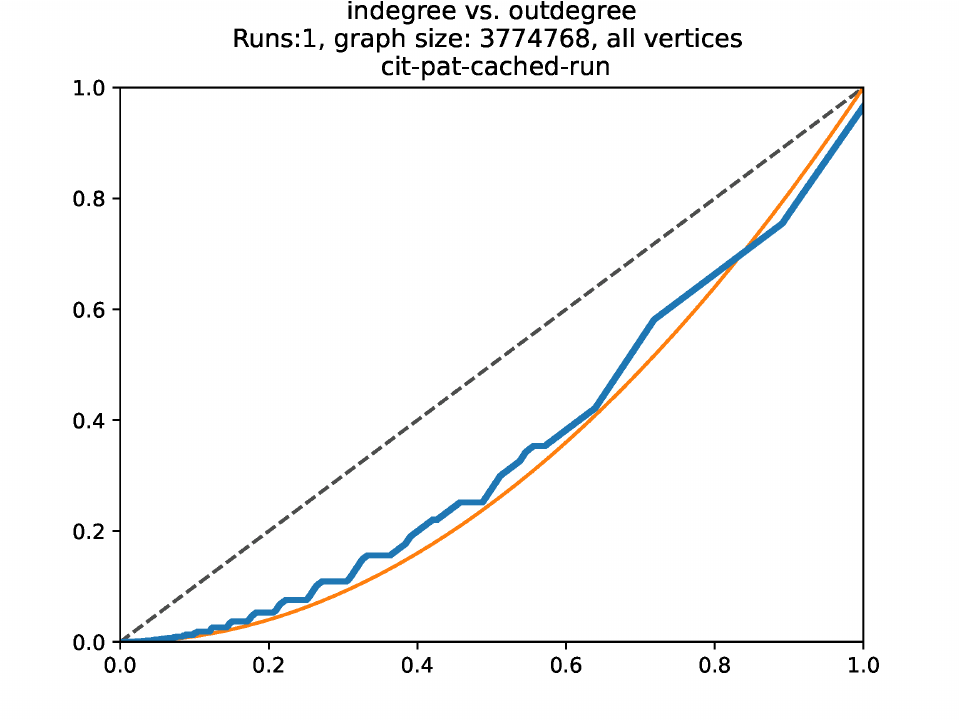}}
\tempoff{\includegraphics[width=0.45\textwidth]{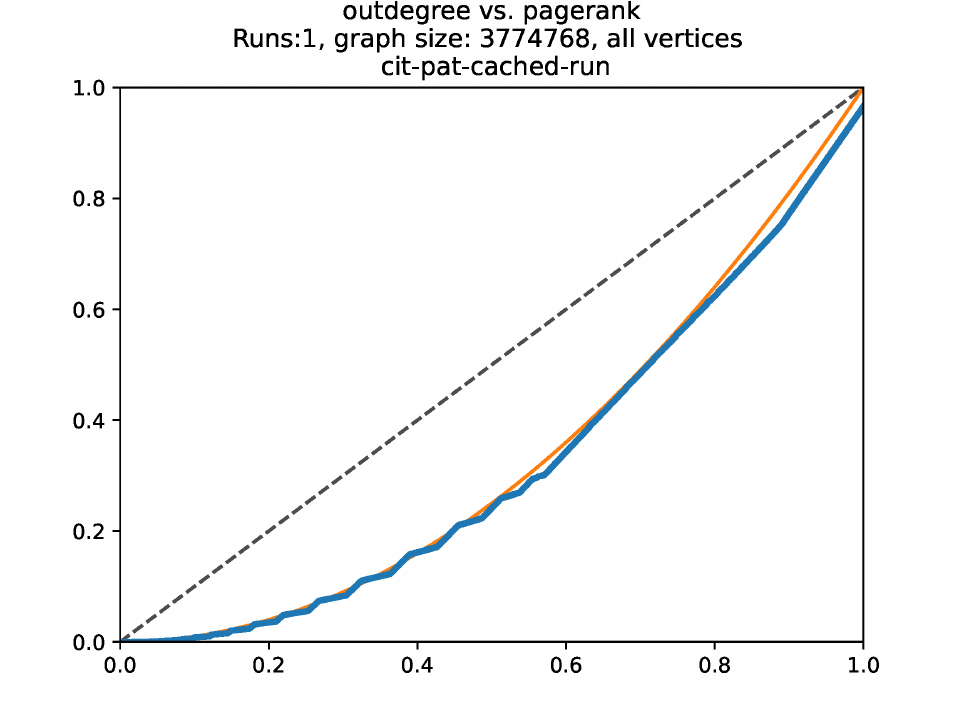}}
\caption{Patent network: CCC for in-degree versus out-degree (left) and out-degree versus PageRank (right).
Both curves are close to $x^2$, indicating near-independence.}
\label{fig:pat-inout-outpr}
\vspace{-0.2cm}
\end{figure}

\noindent

\FloatBarrier

\end{document}